\documentclass[10pt,a4paper]{article}
\usepackage[margin=1in]{geometry}
\usepackage{graphicx}
\usepackage{amsmath}
\usepackage{amssymb}
\usepackage{authblk}
\usepackage[numbers,sort&compress]{natbib}
\usepackage{hyperref}
\newcommand{\Eqref}[1]{Eq.~\ref{#1}}

\title{Field-Modified Quantum Potentials from Tridiagonal Representations: Analytical Spectra and Galerkin Simulations}

\author[1]{Tunde Joseph Osunmusanmi}
\author[1,2,*]{Berihu Teklu}
\affil[1]{College of Computing and Mathematical Sciences, Department of Applied Mathematics and Sciences, Khalifa University of Science and Technology, 127788 Abu Dhabi, United Arab Emirates}
\affil[2]{KU Research Center for Intelligent Computing, Networks and Sensing (ICONS), Khalifa University of Science and Technology, 127788, Abu Dhabi, United Arab Emirates}

\affil[*]{Corresponding author: berihu.gebrehiwot@ku.ac.ae}

\date{\today}

\begin{document}

\maketitle

\begin{abstract}
We develop an effective radial framework for analyzing new field-modified potential function of a charged spinless particle in collinear electric and magnetic fields. Using the weak- magnetic field approximation, we neglect the diamagnetic term and other angular projections under an explicit smallness condition $\ll \Delta E$, where $E$ is the electric field, and the replacement of the angular factor by $C=\cos\theta$ in the spherical coordinates. The resulting equation is therefore a field-aligned radial model whose validity is limited to angularly localized states. Within this setting, we apply the indirect mode (or search mode) of Tridiagonal representation approach in the Laguerre and Jacobi bases. In the Laguerre basis, the tridiagonal constraint generates linear-plus-inverse and linear-plus-quadratic radial potentials, and the coefficient recurrences are connected with Meixner--Pollaczek and Meixner polynomial type structures, therefore yielding the analytic spectra and expansion wavefunctions. In the Jacobi basis, the mapping of the radial half-line to a finite interval produces a singular confining effective potential and a three-term recurrence that is treated through a truncated generalized eigenvalue problem. The formulation provides analytically and numerically tractable benchmark models for field-modified potential functions, while keeping the assumptions, parameter restrictions, endpoint behavior, and convergence tests explicit. In the Laguerre basis the resultant spectra are obtained in closed form and are independent of the orientation and of the electric field, which enter through the shape of the effective central potential, whereas in the Jacobi basis the spectrum is obtained numerically from a new three-term recurrence; in all cases the magnetic field enters through a linear Zeeman shift.
\end{abstract}

\section{Introduction}
\label{intro}
The response of atomic and few-body quantum systems to external electric and magnetic fields is a central problem in atomic, molecular, and optical physics. In hydrogenic systems, electric fields produce Stark mixing of states with different angular momenta, whereas magnetic fields generate Zeeman shifts and, beyond the weak-field limit, diamagnetic confinement. Accurate treatments of these effects use perturbation theory, parabolic-coordinate methods, complex scaling, variational or basis-set diagonalization, and related numerical approaches \cite{BetheSalpeter1957,L D Landau 2013,FauchierDow1974,Osherov VI 2017,Fonte1990,RaoLi1995}. Combined electric and magnetic fields are especially demanding because the exact three-dimensional problem is generally nonseparable; this has motivated extensive studies of crossed, arbitrary-orientation, and parallel-field spectra \cite{Melezhik1993,Main J 1998,JohnsonMotaFurtado2001,Freund2002,Oberreiter2018}.

Here, we propose an effective mathematical  formulation of radial Schrodinger equation in uniform electromagnetic fields using the spherical coordinates where the angular factor is replaced by a parameter $C=\cos\theta$ such that $0<\theta<90$; and retaining the diagonal $m_{\phi}$-dependent energy shift after azimuthal separation. As a result, the equation becomes an effective radial Schr\"{o}dinger equation applicable in centre quantum mechanical problems with uniform electromagnetic fields. Explicitly, it is important to state that the present formulation and the spectra reported below are spectra of the projected radial models, not exact hydrogenic Stark--Zeeman spectra.

Effective radial potentials provide useful reduced descriptions for examining how additional interactions, confinement effects, and external fields may modify bound-state spectra beyond the pure Coulomb problem. In such models, deviations from simple solvable limits are possibly reflected in the eigenvalue distribution, wave-function structure, and recurrence properties of the associated basis expansion. Effective quantum potentials and nonlinear oscillator models have also been used to quantify how departures from harmonicity are reflected in ground-state properties and spectral structure~\cite{Paris2014}. This motivates an algebraic treatment in which field-induced modifications are analyzed through a tridiagonal representation of the radial Hamiltonian.

The method used to analyze the effective radial Schr\"{o}dinger equation  in this work is the Tridiagonal Representation Approach (TRA). The TRA is closely related in spirit to the $J$-matrix program, in which Hamiltonians represented in square-integrable bases lead to Jacobi matrices and three-term recurrence relations for expansion coefficients \cite{HellerYamani1974a,HellerYamani1974b,IsmailKoelink2011,IsmailKoelink2012}. Orthogonal-polynomial and Jacobi-operator theory then provides a natural language for spectra, spectral measures, and finite-truncation approximants \cite{Szego1975,Gautschi2004,Teschl2000,Ismail2005,Koekoek2010}. Recent TRA applications have used this structure to design solvable and quasi-solvable radial quantum systems in Coulomb, Kratzer, and related settings \cite{Alhaidari2017,AlhaidariBahlouli2019,Alhaidari2020CTP,AlhaidariIsmail2023Coulomb,AlhaidariIsmail2023Bessel,AlhaidariLi2025Kratzer,AlhaidariTaiwo2025JMatrix}. Readers are advised to read references on Tridiagonal Representation Approach to have detailed knowledge of this methodology.

In this work, we further extend the application of Tridiagonal representation approach has outlined below. The structure of the manuscript is organized as follows: Section $2$ presents the effective model Hamiltonian and derives the radial Schrödinger equation under the weak-field assumption and angular factor replacement. Section $3$ shows the application of Tridiagonal Representation Approach using the Laguerre basis; and obtain the linear-plus-inverse and linear-plus-quadratic effective radial potentials together with their associated spectra and wavefunctions. Section $4$ extends the construction to a Jacobi basis, where the radial half-line is mapped to a finite interval, and the resulting singular confining potential is treated through a generalized eigenvalue problem. Section $5$ summarizes the main results, discusses the physical interpretation and limitations of the effective radial projection, and outlines possible future extensions. Additional details on the tridiagonal representation method and the relevant orthogonal-polynomial structures are provided in the appendices.

\section{The Effective Model Hamiltonian and Radial Schr\"{o}dinger Equation}

Suppose we consider a nonrelativistic spinless particle of mass $m$ and charge $q$ moving in a central potential $V(r)$ and interacting with static external fields. With minimal coupling, the Hamiltonian is
\begin{equation}
    \hat{H}=\frac{1}{2m}\left(\hat{\mathbf p}-q\mathbf A\right)^2+q\Phi+V(r),
\end{equation}
where $\mathbf A$ and $\Phi$ are vector and scalar potentials. We take uniform electric and magnetic fields parallel to the $z$ axis,
\begin{equation}
    \mathbf B=B\hat{z},\qquad \mathbf E=E\hat{z},
\end{equation}
and use the symmetric gauge and scalar potential
\begin{equation}
    \mathbf A=\frac{1}{2}\left(\mathbf B\times\mathbf r\right),\qquad \Phi(\mathbf r)=-Ez=-Er\cos\theta .
\end{equation}
Expansion of the kinetic term gives the field-free kinetic energy, the linear Zeeman term, and the diamagnetic term $q^2B^2\rho^2/(8m)$, with $\rho=r\sin\theta$. In the weak-magnetic-field regime considered here, the diamagnetic term is neglected when comparably small with respect to the electric field. Hence,
\begin{equation}
    \frac{q^2B^2\rho^2}{8m}\ll \Delta E
\label{eq:T1}
\end{equation}
Retaining the linear magnetic coupling gives
\begin{equation}
    H\simeq \frac{\hat{p}^2}{2m}-\frac{qB}{2m}L_z+V(r)-qEr\cos\theta .
\end{equation}
The electric dipole term $-qEr\cos\theta$ breaks full spherical symmetry and couples angular sectors with $\Delta l=\pm 1$. Therefore, the following is an effective reduction rather than an exact separation. We introduce a parameter
\begin{equation}
    C=\cos\theta,  0\le C\le 1  
\end{equation}
where $\theta$ is a prescribed orientation angle associated with an externally prepared field-aligned configuration. Eventually, $C$ becomes a constant for any value of $\theta$ within $0<\theta<90$; and the angular dependence of the electric field is removed.  Replacing $\cos\theta$ by $C$ yields the effective electric radial term
\begin{equation}
    H_E^{\rm eff}=-qECr .
    \label{eq:TJ}
\end{equation}
The approximation replaces the anisotropic coupling by its projection onto a restricted class of angularly localized state and is valid only within the assumptions stated above. So the neglected component of the field projection is $\Delta \delta=-qEr\left( \cos\theta -C \right)$ similiar to Eq.~\ref{eq:T1}, can also be removed as long as $|qE|r_{typ}\sqrt{Var\left( \cos\theta \right)}\ll \Delta E$. 
After the projection, the azimuthal dependence $e^{i m_{\phi}\phi}$ diagonalizes the linear magnetic term. The effective central equation may therefore be written as
\begin{equation}
    \left[-\frac{\hbar^2}{2m}\nabla^2+V(r)-qECr\right]\Psi(r,\theta,\phi)=\acute{\xi}\,\Psi(r,\theta,\phi),
    \label{eq:A}
\end{equation}
with the resultant energy spectrum (caused as result of the linear magnetic term)
\begin{equation}
    \acute{\xi}=\xi+\frac{qB}{2m}m_{\phi}\hbar 
\end{equation}
where $\xi$ represents the energy spectrum. Writing $\Psi(r,\theta,\phi)=R_{nl}(r)Y_{lm_{\phi}}(\theta,\phi)$ and defining $\psi(r)=rR_{nl}(r)$ gives the reduced one-dimensional radial equation
\begin{equation}
    \frac{d^2\psi}{dr^2}+\left[\frac{2m}{\hbar^2}\left(\acute{\xi}-\acute{V}(r)\right)-\frac{l(l+1)}{r^2}\right]\psi(r)=0,
    \label{eq:B}
\end{equation}
where
\begin{equation}
    \acute{V}(r)=V(r)-qECr.
\end{equation}
Since $V(r)=\acute{V}(r)+qECr$, the quantity $V(r)$ is the effective field modified central radial potential to be determined by the tridiagonal constraint, which is made up of the sum of $\acute{V}(r)$, the missing  effective radial potential part and $qECr$ the effective electric term. The parameter $C$ of the potential function $V(r)$ can be seen as a variable that changes to give different structural form to $V(r)$ as $\theta$ varies. So for each $\theta$, $V(r)$ is well defined and as a result a combined plot of $V(r)$ can be obtained when $C$ is varied. The total effective field-modified central potential will be the addition of $V(r)$ and the centrifugal term $\frac{l(l+1)}{2r^2}$. For simple clarity, the proposed effective radial Schrodinger equation is
\begin{equation}
    \left[ -\frac{\hbar^2}{2m}\frac{d^2}{dr^2}+ \frac{\hbar^2}{2m}\frac{l(l+1)}{r^2}+ \acute{V}(r) \right]\psi(r)=\acute{\xi}\psi(r)
    \label{eq:Z1}
\end{equation}
where $\acute{V}(r)=V(r)-qECr$ is the potential that enters the radial operator, $V(r)=\acute{V}(r)+qECr$ is the effective potential function to be determined, $\xi$ is the energy spectrum of the physical system, and $\acute{\xi}$ is the sum or resultant of the energy spectrum $\xi$ and linear magnetic term $\frac{qB}{2m}m_{\phi}\hbar$. Invariably, it means $\acute{\xi}_{k,l,m_{\phi}}=\xi_{k}+\frac{qBm_{\phi}\hbar}{2m}$. Since the radial operator contains only $\acute{V}(r)$, the resultant spectrum $\acute{\xi}$ is fixed by $\acute{V}(r)$, while the orientation and the electric field enter through the shape of $V(r)$. Accordingly, in the figures below the energy levels are displayed either together with the potential they belong to, $\acute{V}(r)+\frac{l(l+1)}{2r^{2}}$, or together with $V(r)$ for several orientations; in all cases they are eigenvalues of the radial operator containing $\acute{V}(r)$. Henceforth, we generally refer to $V(r)$ as the effective field-modified central potential, or simply the central potential. The Hamiltonian in Eq~\ref{eq:Z1} provides an effective radial description of a charged spinless particle subjected to parallel electric and magnetic fields under the weak-field and angular-localization assumptions introduced above. It is not intended to represent the exact three-dimensional Stark–Zeeman Hamiltonian, but rather a reduced model that captures the dominant field-induced modifications of the radial dynamics. Such effective Hamiltonians are valuable because they preserve the self-adjoint radial Schrödinger structure while remaining analytically tractable. Consequently, they serve as benchmark models for investigating how external fields modify central potentials, bound-state spectra, wavefunctions, and confinement properties. Furthermore, the radial form is particularly well suited for the Tridiagonal Representation Approach, enabling the systematic construction of exactly solvable field-modified potentials and their associated orthogonal-polynomial solutions. 

\section{TRA Solution of the Effective Radial Schr\"{o}dinger Equation in Laguerre Basis}
\subsection{Linear plus inverse radial potential}

We consider the solution of Eq~\ref{eq:B}, using the Laguerre basis \cite{Magnus W 1966,Ismail2005} with boundary condition $\psi(0)=0$ and $\psi(r)=0$ as $r\to \infty $. With this, we analytically obtain the central potential function, energy spectrum, and the wavefunction. In atomic unit of $\hbar= m = 1$, Eq~\ref{eq:B} becomes 
\begin{equation}
    \left( H-\acute{\xi} \right)\psi(r)=\left[ \frac{-1}{2}\frac{d^2}{dr^2}+\frac{l(l+1)}{2r^2} +\acute{V}(r)-\acute{\xi}\right]\psi(r)=0
\label{eq:G}
\end{equation}
where $V\acute{}=V(r)-qECr$, $\acute{\xi}=\xi+\frac{qBm_{\phi}}{2}$, $m_{\phi}$ is the azimuth quantum number, $E$  and $B$  are electric and magnetic fields, $\acute{\xi}$ is the resultant energy spectrum. Using space transformation $x = \lambda r$, where $\lambda$ is a positive dimensionless scale parameter having an inverse length, and with the basis element 
\begin{equation}
    \phi_{n}(r)=A_{n}x^\alpha e^{-\beta x}L^{v}_{n}(x)
\label{eq:H}
\end{equation}
where $A_{n}=\sqrt{\frac{\lambda \Gamma (n+1)}{\Gamma(n+\nu+1)}}$ , $\alpha$ and $\beta$ are the basis parameters, and $L^{\nu}_{n} (x)$ is a Laguerre polynomial such that $\nu>-1$ and $n=0,1,2,...,$. Using this basis, Eq~\ref{eq:G} becomes 
\begin{equation}
\label{eq:long_equation}
\begin{aligned}
\frac{-2}{\lambda^2}
\left( H-\acute{\xi} \right)\phi_{n}(r)
&= A_{n}x^{\alpha}e^{-\beta x}
\biggl[
\biggl\{
-\frac{\nu+1-x}{2}
-\left( \frac{2\alpha}{x}-2\beta \right)
\biggr\}
\frac{d}{dx}
\\
&\quad
+\frac{\alpha(\alpha-1)-l(l+1)}{x^2}
-\frac{2\left(V(x)-\acute{\xi}\right)}{\lambda^2}
\\
&\quad
-\frac{n+2\alpha\beta}{x}
+\beta^2
+\frac{2qECx}{\lambda^3}
\biggr]
L^{\nu}_{n}(x).
\end{aligned}
\end{equation}
Taking the basis parameter as $2\alpha=\nu+1$, $\beta=\frac{1}{2}$, and $\alpha(\alpha-1)=l(l+1)$, we have
\begin{equation}
\label{eq:E}
\begin{aligned}
\frac{2}{\lambda^2}
(H-\acute{\xi})\phi_n(r)
&=A_{n}x^{\alpha-1}e^{-\beta x}
\biggl[
(n+\alpha)
-\frac{x}{4}
\\
&\quad
+\frac{2x}{\lambda^2}
\bigl(V(x)-\acute{\xi}\bigr)
-\frac{2qECx^2}{\lambda^3}
\biggr]
L^{\nu}_{n}(x).
\end{aligned}
\end{equation}
At this stage we are employing the search mode of the Tridiagonal Representation Approach. The purpose is to determine the effective potential for which the wave operator admits a tridiagonal representation in the chosen basis. Consequently, the parameterization introduced below should be regarded as an algebraic constraint imposed by the tridiagonality requirement ensuring that the Hamiltonian parameters do not depend on the energy spectrum. These  constants are independent model parameters used to identify separately the effective potential $V(r)$ and the resultant energy spectrum $\acute{\xi}$. Therefore,
\begin{equation}
    \frac{2x}{\lambda^2}\left( V(x)-\acute{\xi}\right)=\frac{2qECx^2}{\lambda^3}+u_{1}x+u_{0}
\label{eq:C}
\end{equation}
From Eq~\ref{eq:G}, as an energy-independent potential function, we have 
\begin{equation}
    V(r)=ar+\frac{b}{r},
\label{eq:K}
\end{equation}
where $a=qEC$ and $b=\frac{\lambda u_{0}}{2}$. Correspondingly, the potential entering the radial operator is $\acute{V}(r)=V(r)-qECr=b/r$.
\begin{figure}
    \centering
    \includegraphics[width=0.72\linewidth]{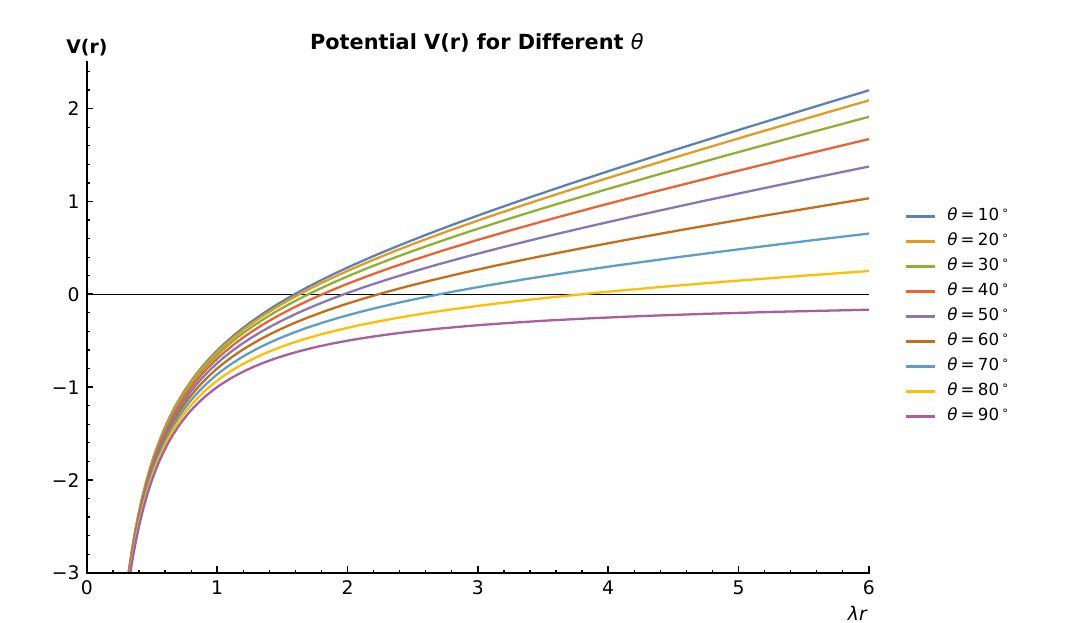}
    \caption{Graph for the potential function $V(r)$ Eq~\ref{eq:K} with $\lambda=1$, $E=2.0$, $q=0.2$, $C=\cos\theta$, and $u_{0}=-2$ ($b=-1$). For $b<0$, $V(r)$ increases monotonically from $-\infty$ at the origin, and the orientation controls the slope of the linear outer wall. The curve $\theta=90^{\circ}$ ($C=0$) coincides with $\acute{V}(r)=b/r$. The vertical axis is in a dimensionless scaled energy unit, and the horizontal axis is in a dimensionless radial coordinate unit.}
    \label{fig:1}
\end{figure}
$V(r)$ is a central-type effective potential consisting of two competing contributions - A linear term proportional to $r$ and an inverse-distance term proportional to $1/r$. For $a>0$ and $b>0$ the derivative of $V(r)$ vanishes at $r_{min}=\sqrt{b/a}$, where $V(r)$ has a minimum. However, the radial operator in Eq.~\ref{eq:G} contains only $\acute{V}(r)=b/r$, which is repulsive for $b>0$ and supports no bound states. Bound states require an attractive component, $b<0$ (that is, $u_{0}<0$). In this case $V'(r)=a+|b|/r^{2}>0$, so that $V(r)$ increases monotonically from $-\infty$ at the origin; the Coulomb-like attraction dominates at short distances, and the linear term forms an outer wall whose slope is controlled by the orientation through $C=\cos\theta$, as shown in Fig.~\ref{fig:1}. The absence of an interior minimum in $V(r)$ does not preclude bound states: the radial operator contains $\acute{V}(r)=-|b|/r$, an attractive Coulomb-like potential which, as in the hydrogenic problem, supports an infinite discrete spectrum without possessing a minimum. For $l\ge1$ the centrifugal term produces the familiar well $-|b|/r+l(l+1)/2r^{2}$, with minimum at $r=l(l+1)/|b|$ and depth $-b^{2}/[2l(l+1)]$ (Fig.~\ref{fig:3}). The binding is thus controlled by the sign of $b$, i.e. of $u_{0}$, and not by the shape of $V(r)$. Also, from Eq~\ref{eq:C}, it therefore necessitates that \begin{equation}
    \acute{\xi}=\frac{-\lambda^2 u_{1}}{2}
\label{eq:D}
\end{equation}
where $u_{0}$ and $u_{1}$ are real parameters with distinct roles. The parameter $u_{0}$ defines the potential $V(r)$ in Eq.~\ref{eq:K} through $b=\lambda u_{0}/2$; once $u_{0}$ and $\lambda$ are fixed, the Hamiltonian is a single well-defined operator that does not vary from one eigenstate to another. The parameter $u_{1}$, on the other hand, carries the resultant energy through Eq.~\ref{eq:D}. Eq.~\ref{eq:D} should therefore be read not as a formula for the spectrum, but as the identification, introduced through the parameterization of Eq.~\ref{eq:C}, that reduces the differential equation to a three-term recursion (tridiagonal) form in which the energy appears only through $u_{1}$. The discrete values $\acute{\xi}_{k}$ that are admissible for $u_{1}$ are then obtained from the resulting three-term recursion, Eq.~\ref{eq:T2}, exactly as in a standard eigenvalue problem, and the physical spectrum follows as $\xi_{k}=\acute{\xi}_{k}-\frac{qB}{2}m_{\phi}$. Therefore, the energy level shift becomes $\Delta\xi=\acute{\xi}_{k}-\xi_{k}=\frac{qB}{2}m_{\phi}$, which is directly proportional to the linear magnetic term and independent of $k$ and $l$. The above explanations are applicable to the subsequent subsection, where the roles of the potential and energy parameters are played by $n_{1}$ and $n_{0}$, respectively.

Therefore, Eq~\ref{eq:E} will become
\begin{equation}
    \frac{2}{\lambda^2}\left( H-\acute{\xi} \right)\phi_{n}(r) = A_{n}x^{\alpha-1}e^{-\beta x}\left[ (n+\alpha+u_{0})+u_{2}\,x \right] L_{n}^{\nu}(x),\qquad u_{2}=u_{1}-\frac{1}{4}.
\end{equation}
Using the recursion relation and orthogonality properties of the Laguerre polynomial, the wave equation becomes an  energy polynomial with three term recursion
\begin{equation}
\begin{aligned}
\Bigl[(2u_{2}+1)(n+\mu)+u_{0}\Bigr]f_{n}
&=u_{2}\Bigl[\sqrt{n(n+\nu)}\,f_{n-1}
\\
&\quad
+\sqrt{(n+1)(n+\nu+1)}\,f_{n+1}\Bigr],
\end{aligned}
\label{eq:T2}
\end{equation}
where $\mu=\frac{\nu+1}{2}=l+1$, since $2\alpha=\nu+1$ and $\alpha=l+1$. In Eq.~\ref{eq:T2} the potential parameter $u_{0}$ and the energy parameter $u_{2}$ play different roles, and the diagonal term depends on the basis index $n$ only. The recursion coefficients are independent of the spectral label $k$, which enters solely through the admissible values of $u_{2}$. Comparing Eq.~\ref{eq:T2} with the standard Meixner--Pollaczek three-term recurrence (B8) in Appendix B, written as $\frac{1}{2}\bigl[\sqrt{n(n+2\mu-1)}\,P_{n-1}+\sqrt{(n+1)(n+2\mu)}\,P_{n+1}\bigr]=\bigl[z\sin\varphi+(n+\mu)\cos\varphi\bigr]P_{n}$, one observes the same tridiagonal structure with
\begin{equation}
    \mu=\frac{\nu+1}{2},\qquad \cos\varphi=\frac{2u_{2}+1}{2u_{2}},\qquad z\sin\varphi=\frac{u_{0}}{2u_{2}}.
\label{eq:F}
\end{equation}
From Eq.~\ref{eq:F}, $\cos^{2}\varphi-1=u_{1}/u_{2}^{2}$, so that $|\cos\varphi|>1$ whenever $\acute{\xi}<0$, i.e. $u_{1}>0$. The bound-state sector therefore lies outside the continuous range of the Meixner--Pollaczek polynomial and is reached by the analytic continuation $\varphi\to i\vartheta$, with $\sinh^{2}\vartheta=u_{1}/u_{2}^{2}$. Working with the bound state energy spectrum obtained from the pole in the amplitude of the asymptotics of the Meixner--Pollaczek polynomial, $\mu+iz=-k$, one obtains $(k+\mu)^{2}u_{1}=u_{0}^{2}/4$, and the resultant energy spectrum of the system becomes
\begin{equation}
    \acute{\xi}_{k}=-\frac{\lambda^{2}u_{0}^{2}}{8(k+l+1)^{2}}=-\frac{b^{2}}{2(k+l+1)^{2}},\qquad
    \xi_{k,l,m_{\phi}}=\acute{\xi}_{k}-\frac{qB}{2}m_{\phi},
\label{eq:L}
\end{equation}
for $k=0,1,2,3,\ldots$, with bound states only for $b<0$. The spectrum is independent of the basis scale $\lambda$, as it must be for a fixed Hamiltonian, and, because the electric term $qECr$ is contained in $V(r)$ and not in $\acute{V}(r)$, it is also independent of the orientation $\theta$ and of the electric field $E$; these enter through the shape of $V(r)$ shown in Fig.~\ref{fig:1}.
For each level, the analytic continuation $\varphi\to i\vartheta$ maps the Meixner--Pollaczek recurrence onto the discrete Meixner recurrence (B14) with parameter $\beta_{k}=e^{-2\vartheta_{k}}=w_{k}^{2}$, where
\begin{equation}
    w_{k}=\frac{|u_{0}|-(k+\mu)}{|u_{0}|+(k+\mu)}=\frac{2|b|-\lambda(k+l+1)}{2|b|+\lambda(k+l+1)},
\label{eq:wk}
\end{equation}
so that $0\le\beta_{k}<1$. Hence, $f_{n}(\acute{\xi}_{k})\propto s_{k}^{\,n}M_{n}^{\mu}(k;\beta_{k})$ with $s_{k}=\operatorname{sgn}(w_{k})$. For $\lambda=2|b|/(k+l+1)$ one has $\beta_{k}=0$, and the expansion reduces to the single basis element $\phi_{k}$.

\begin{figure}
    \centering
    \includegraphics[width=0.7\linewidth]{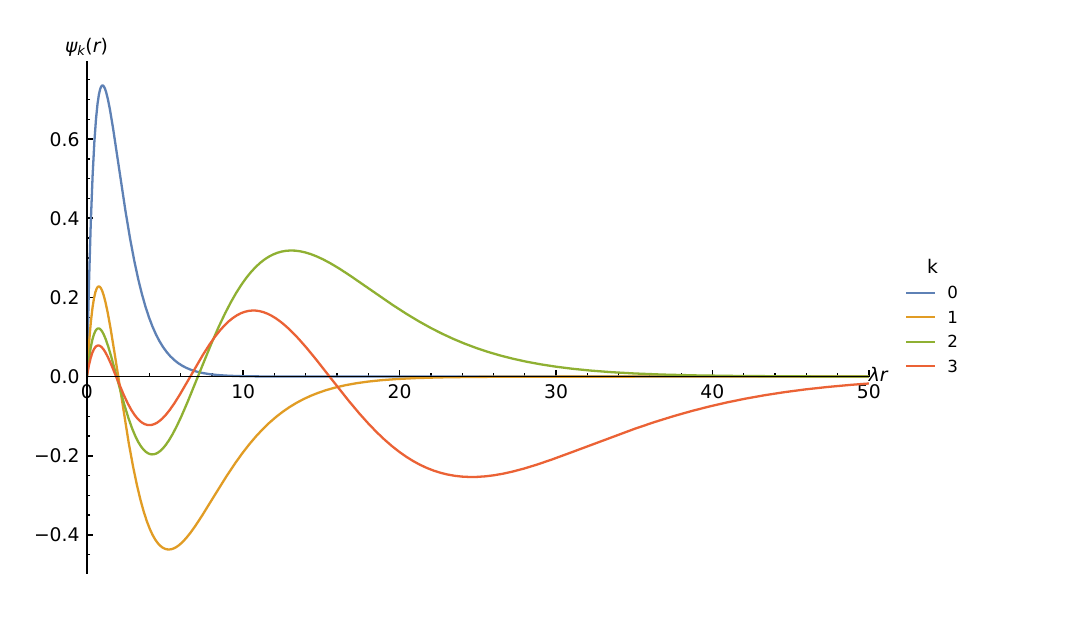}
    \caption{Graph of the normalized wavefunction $\psi_{k}\left( r\right)$, Eqs.~\ref{eq:J} and \ref{eq:Jc}, for the localized states $l=0$ at each energy level $\acute{\xi}_{k}$ Eq~\ref{eq:L}, where $k=0,1,2,3$ and $b=-1$. The state $k$ has $k$ nodes, and the wavefunctions are independent of $\theta$. The vertical axis is dimensionless scaled wavefunction amplitude, and the horizontal axis is in a dimensionless radial coordinate unit.}
    \label{fig:4}
\end{figure}

\begin{figure}
    \centering
    \includegraphics[width=0.7\linewidth]{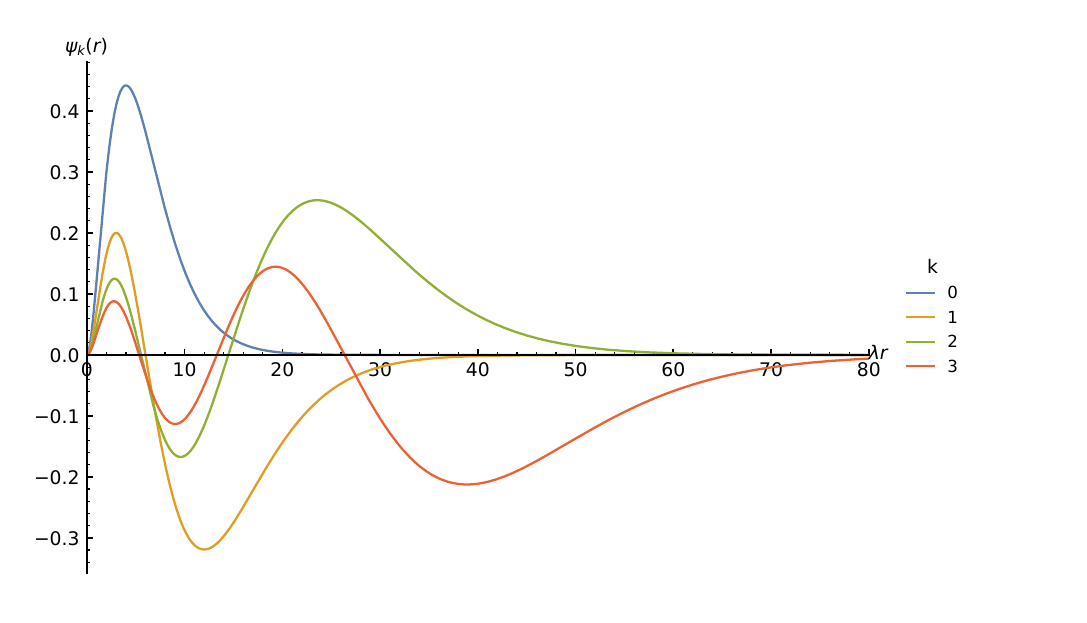}
    \caption{ Graph of the normalized wavefunction $\psi_{k}\left(r\right)$, Eqs.~\ref{eq:J} and \ref{eq:Jc}, for the first angularly excited p-state or dipole like state $l=1$ at each energy level $\acute{\xi}_{k}$ Eq~\ref{eq:L}, where $k=0,1,2,3$ and $b=-1$. The vertical axis is dimensionless scaled wavefunction amplitude, and the horizontal axis is in a dimensionless radial coordinate unit.}
    \label{fig:5}
\end{figure}

Therefore, the bound-state wavefunction is written in terms of the Meixner polynomial, which is the discrete version of the Meixner-Pollaczek polynomial with an infinite spectrum
\begin{equation}
    \psi\left( \acute{\xi}_{k},r \right)=\frac{\sqrt{|b|\,\rho_{k}(\beta_{k})}}{k+l+1}\sum_{n=0}^{\infty}s_{k}^{\,n}M_{n}^{\mu}(k;\beta_{k})\,\phi_{n}(\lambda r)
\label{eq:J}
\end{equation}
where $M_{n}^{\mu}(k;\beta)=\sqrt{\frac{\Gamma(n+2\mu)}{\Gamma(2\mu)\Gamma(n+1)}}\,\beta^{n/2}{}_2F_1\left(
\begin{array}{c}
-n,-k \\
2\mu
\end{array}
\middle| 1-\beta^{-1}
\right)$
and $\rho_{k}(\beta)=\beta^{k}(1-\beta)^{2\mu}\frac{\Gamma(k+2\mu)}{\Gamma(2\mu)\Gamma(k+1)}$ is the Meixner weight. The prefactor normalizes $\psi$ to unity; it differs from $\sqrt{\rho_{k}}$ because, for $x=\lambda r$, the basis of Eq.~\ref{eq:H} is not orthonormal, its overlap matrix being the tridiagonal matrix generated by $xL_{n}^{\nu}(x)$. Summing the series in Eq.~\ref{eq:J} gives the closed form
\begin{equation}
    \psi_{k}(r)=N_{k}\,r^{l+1}e^{-c_{k}r}L_{k}^{2l+1}(2c_{k}r),\qquad c_{k}=\frac{|b|}{k+l+1},\qquad N_{k}^{2}=\frac{(2c_{k})^{2l+3}\,k!}{2(k+l+1)\,\Gamma(k+2l+2)},
\label{eq:Jc}
\end{equation}
which is used in Figs.~\ref{fig:4} and \ref{fig:5}. 

Eq.~\ref{eq:J} is well defined eigenfunction with analytical properties such as convergence, square integrability, satisfaction of the boundary conditions are well established in foundational TRA references. Readers are therefore advised to read these references. Here we give a short overview. The basis functions $\phi_{n}(\lambda r)$ defined in Eq.~\ref{eq:H} with parameters $\alpha,\beta\ge0$ and $\nu>-1$, form a complete set in $L^{2}(0,\infty)$, and also each individual $\phi_{n}(r)$ already satisfies the required boundary conditions $\phi_{n}(0)=0$ and $\phi_{n}(r)\to0$ as $r\to\infty$. The expansion coefficients $M_{n}^{\mu}(k;\beta_{k})$ belong to the discrete Meixner family with parameter $0\le\beta_{k}<1$ as said earlier; these coefficients are square-summable with their orthogonality weight by construction of the Meixner orthogonality relation, so the series in Eq.~\ref{eq:J} converges in the $L^{2}(0,\infty)$ norm to the closed form of Eq.~\ref{eq:Jc}. In addition, because the coefficients decay geometrically in $n$ (as $|w_{k}|^{n}$, up to a slowly varying ratio of Gamma functions), the series also converges absolutely and uniformly on every compact subset of $(0,\infty)$. Uniform convergence on compacts, together with the fact that every partial sum satisfies the boundary conditions, guarantees that the limit function $\psi(\acute{\xi}_{k},r)$ inherits $\psi(0)=0$ and $\psi(r)\to0$ as $r\to\infty$. Since $H|\phi^{\lambda}_{n}\rangle=a_{n}|\phi^{\lambda}_{n}\rangle+b_{n-1}|\phi^{\lambda}_{n-1}\rangle+b_{n}|\phi^{\lambda}_{n+1}\rangle$, then $\psi(\acute{\xi}_{k},r)$ is in the domain of the associated self-adjoint radial operator. All analysis and discussions held in this section are applicable in the subsequent ones. 

\begin{figure}
    \centering
    \includegraphics[width=0.7\linewidth]{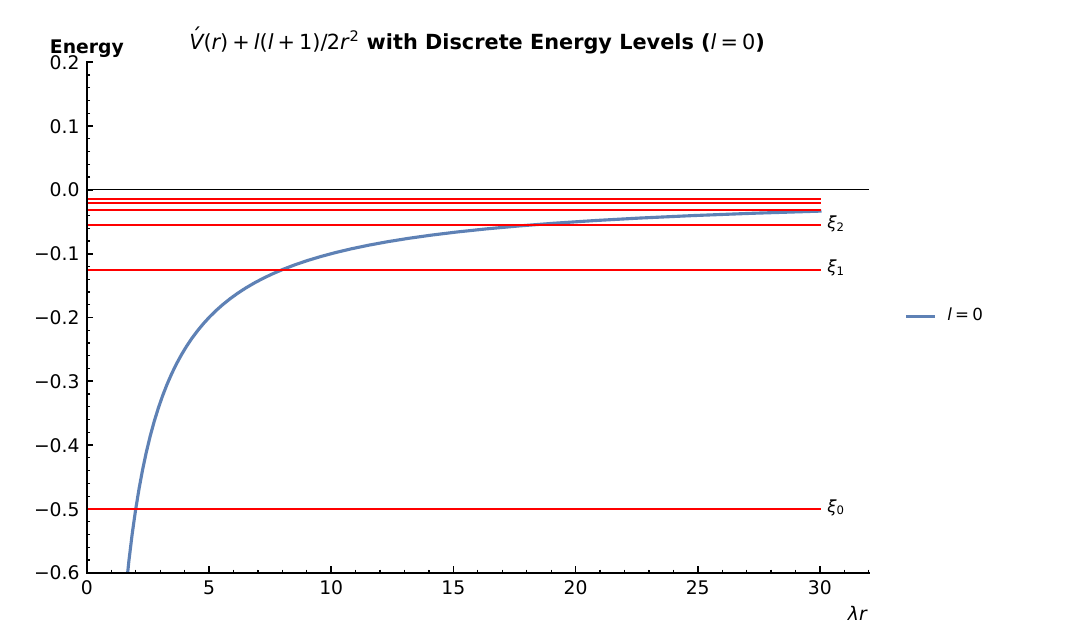}
    \caption{Graph of the potential entering the radial operator, $\acute{V}(r)+\frac{l(l+1)}{2r^{2}}$ with $\acute{V}(r)=b/r$, and the $\theta$-independent resultant energy spectrum $\acute{\xi}_{k}=-b^{2}/[2(k+l+1)^{2}]$ Eq~\ref{eq:L}, $k=0,\ldots,5$, for $l=0$ ($\nu=1$), $\lambda=1$ and $b=-1$. The levels accumulate at $\acute{\xi}=0^{-}$. The vertical axis is in a dimensionless scaled energy unit, and the horizontal axis is in a dimensionless radial coordinate unit.}
    \label{fig:2}
\end{figure}
\begin{figure}
    \centering
    \includegraphics[width=0.7\linewidth]{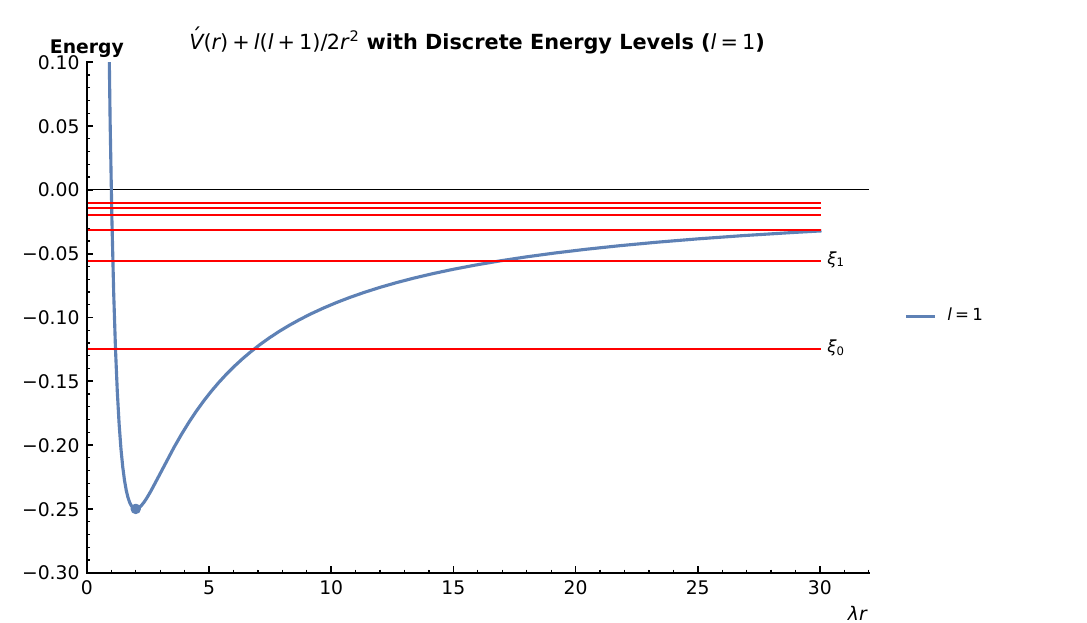}
    \caption{As Fig.~\ref{fig:2} for $l=1$ ($\nu=3$). The centrifugal term produces a well with minimum at $r=l(l+1)/|b|=2$ (dot) and depth $-0.25$, which contains the ground level $\acute{\xi}_{0}=-0.125$. The vertical axis is in a dimensionless scaled energy unit, and the horizontal axis is in a dimensionless radial coordinate unit.}
    \label{fig:3}
\end{figure}

\subsection{Quadratic plus Linear radial potential}
\begin{figure}
    \centering
    \includegraphics[width=0.7\linewidth]{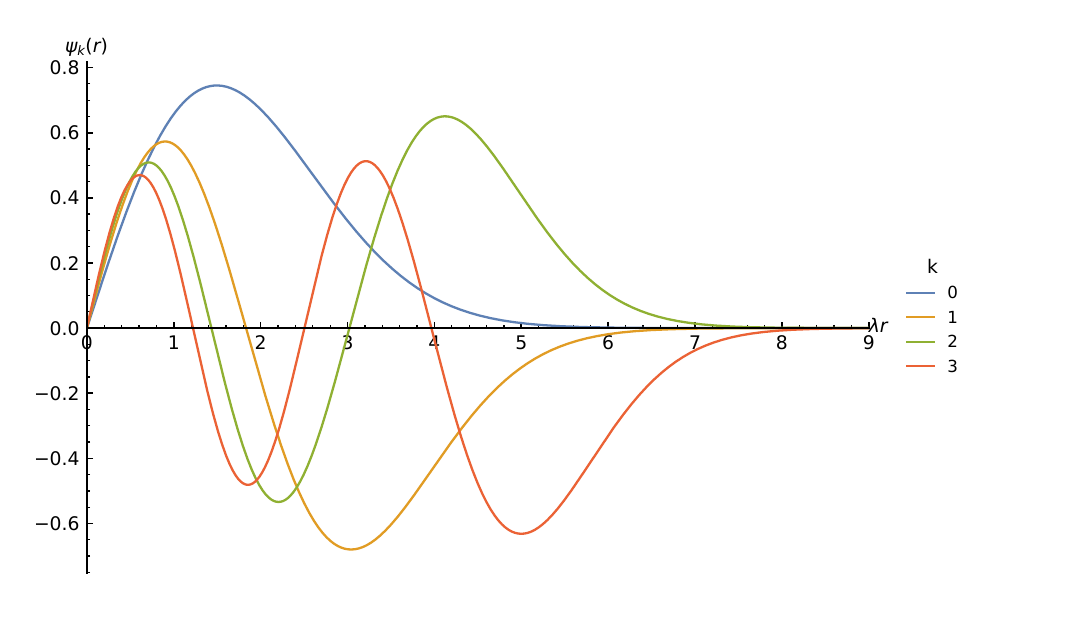}
    \caption{Graph of the normalized wavefunction $\psi_{k}\left( r \right)$, Eq.~\ref{eq:J2}, for the localized states $l=0$ at each energy level $\acute{\xi}_{k}$ Eq~\ref{eq:O}, where $k=0,1,2,3$ and $\eta=0.1$. The wavefunctions are independent of $\theta$. The vertical axis is dimensionless scaled wavefunction amplitude, and the horizontal axis is in a dimensionless radial coordinate unit.}
    \label{fig:8}
    
\end{figure}

\begin{figure}
    \centering
    \includegraphics[width=0.7\linewidth]{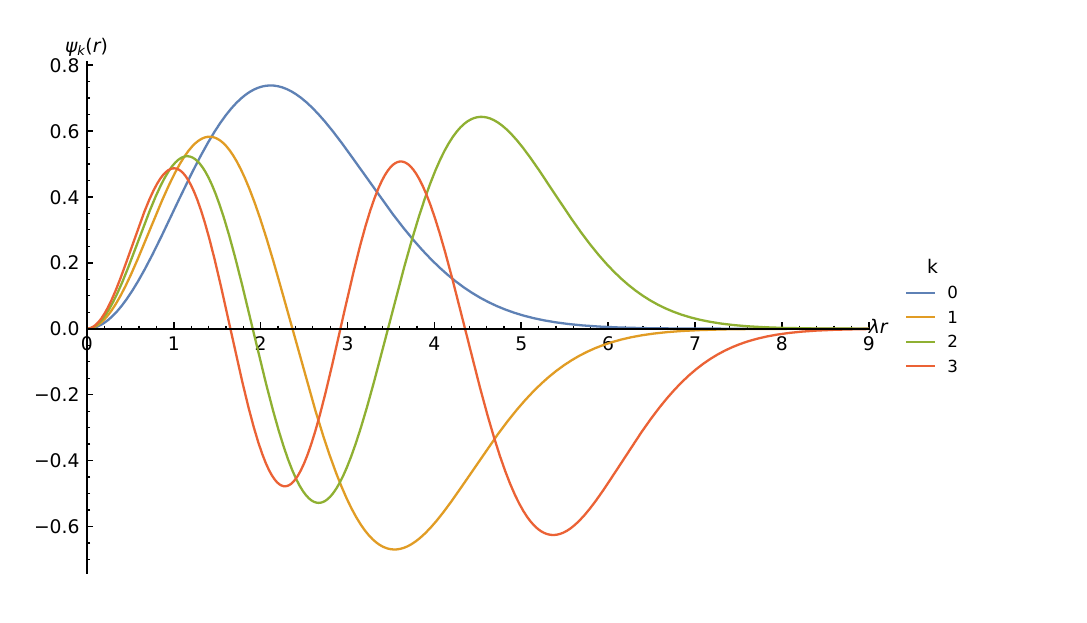}
    \caption{ Graph of the normalized wavefunction $\psi_{k}\left( r \right)$, Eq.~\ref{eq:J2}, for the dipole like states or first angularly excited p-state $l=1$ at each energy level $\acute{\xi}_{k}$ Eq~\ref{eq:O}, where $k=0,1,2,3$ and $\eta=0.1$. The vertical axis is dimensionless scaled wavefunction amplitude, and the horizontal axis is in a dimensionless radial coordinate unit.}
    \label{fig:9}
\end{figure}

From Eq~\ref{eq:G}, using the same basis element Eq~\ref{eq:H}, but with a normalization factor $A_{n}=\sqrt{\frac{2\lambda\Gamma(n+1)}{\Gamma(n+\nu+1)}}$ and a space transformation $x=\lambda^2r^2$, gives
\begin{equation}
\label{eq:split_equation}
\begin{aligned}
\frac{-1}{2\lambda^2}
\left(H-\acute{\xi}\right)\phi_n(r)
&=
A_n x^{\alpha} e^{-\beta x}
\biggl\{
\Bigl[
-\nu-\frac12+2\alpha
+x(1-2\beta)
\Bigr]
\frac{d}{dx}
\\
&\quad
+\frac{\alpha^2-\frac{\alpha}{2}-\frac{l(l+1)}{4}}{x}
-2\alpha\beta
+x\beta^2
\\
&\quad
-\frac{\beta}{2}
-n
-\frac{1}{2\lambda^2}
\bigl(V(x)-\acute{\xi}\bigr)
\\
&\quad
+\frac{qEC}{2\lambda^3}\sqrt{x}
\biggr\}
L_n^{\nu}(x).
\end{aligned}
\end{equation}
Taking $\beta=\frac{1}{2}$, $2\alpha=\nu+\frac{1}{2}$, and $\alpha^2-\frac{\alpha}{2}=\frac{l(l+1)}{4}$ gives $\alpha=\frac{l+1}{2}$ (the root regular at the origin) and $\nu=l+\frac{1}{2}$; therefore, we have 
\begin{equation}
\label{eq:N}
\begin{aligned}
\frac{1}{2\lambda^2}
\left(H-\acute{\xi}\right)\phi_n(r)
&=
A_n x^{\alpha} e^{-\beta x}
\biggl\{
\frac{4n+4\alpha+1}{4}
-\frac{x}{4}
\\
&\quad
+\frac{1}{2\lambda^2}
\bigl(V(x)-\acute{\xi}\bigr)
-\frac{qEC}{2\lambda^3}\sqrt{x}
\biggr\}L_n^{\nu}(x).
\end{aligned}
\end{equation}
Similar to the first case, our objective is to determine the effective potential that admit a tridiagonal representation in the prescribed basis. The parameterization introduced below is therefore an algebraic consequence of the tridiagonality condition and should not be interpreted as implying an energy-dependent Hamiltonian. Rather, the auxiliary constants are independent model parameters introduced to distinguish explicitly between the effective potential $V(r)$ and the resultant energy $\acute{\xi}$.

\begin{equation}
    \frac{1}{2\lambda^{2}}\left(V(x)-\acute{\xi}\right)=\frac{qEC}{2\lambda^{3}}\sqrt{x}+n_{1}x+n_{0}
\label{eq:I}
\end{equation}
where $n_{1}$ (potential parameter) and $n_{0}$ (energy parameter) are real parameters, with roles analogous to $u_{0}$ and $u_{1}$ in the previous subsection. From Eq~\ref{eq:I}, we have
\begin{equation}
    V(r)= \kappa r+\eta r^2
\label{eq:M}
\end{equation}
where $\kappa = qEC$ and $\eta = 2n_{1}\lambda^4$. This represents a superposition of a linear and a quadratic radial dependence. Both terms of the potential describe a confined system subjected to a constant external field, a configuration that arises naturally in a variety of physical contexts. A key aspect of this potential is its ability to support bound states. The existence of bound states is governed by the asymptotic behavior of the potential at large distances. As the linear term grows only linearly with $r$, the quadratic term dominates for sufficiently large $r$. Consequently, the potential behaves as $V(r)\approx \eta r^2$. For the system to remain stable and admit bound states, it is therefore necessary that the coefficient of the quadratic term be positive $\eta$, i.e., $\eta > 0$. Under this condition, the potential increases without bound at large distances, ensuring spatial confinement of the particle and the existence of a discrete energy spectrum. If, on the other hand, $\eta < 0$, the potential becomes unbounded from below, leading to an unstable system with no physically meaningful bound states. The minimum value of the potential is lowered to 
\begin{equation}
    r_{min}=max\left( 0,-\frac{\kappa}{2\eta} \right)
\end{equation}
as $r\ge 0$.  Physically, this corresponds to a balance between the restoring force of the isotropic quadratic confinement and the constant force exerted by the electric field.  However, the quantization of energy levels are preserved. The discreteness of the spectrum follows from confinement, which can be demonstrated analytically or verified numerically. Potentials of this form are widely encountered in several areas of modern physics. In semiconductor physics, they provide a useful model for electrons confined in quantum dots under external electric fields, which can be used to manipulate the spatial localization of charge carriers. In atomic and molecular physics, the potential serves as a simplified model for the Stark effect, describing how external electric fields perturb bound states. In the context of trapped-ion and cold-atom systems, harmonic confinement combined with external fields is routinely employed for precision control and quantum information processing. More broadly, such potentials are relevant in any system where a confining mechanism coexists with a uniform external force, making them fundamental and versatile tools in both theoretical and applied physics.
\begin{figure}
    \centering
    \includegraphics[width=0.7\linewidth]{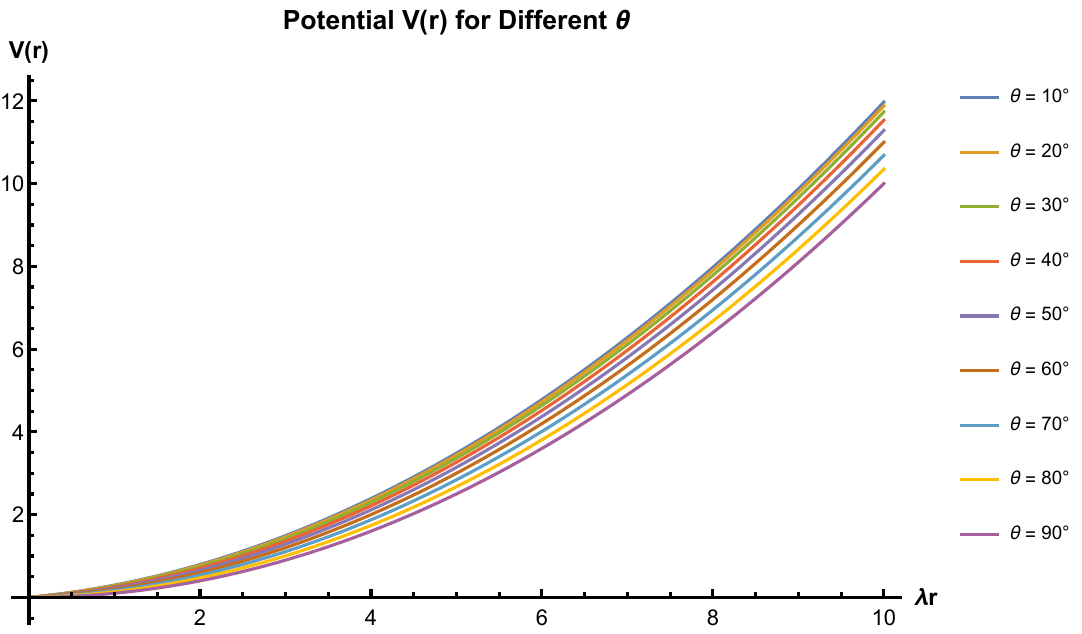}
    \caption{Graph of the potential function $V(r)$ Eq~\ref{eq:M} with $\kappa = 0.2 \cos\theta$ and $\eta = 2n_{1} \lambda^{4}$ where $\lambda=1$, $E=1$, $q=0.2$, and $n_{1}=0.05$ ($\eta=0.1$). The vertical axis is in a dimensionless scaled energy unit, and the horizontal axis is in a dimensionless radial coordinate unit.}
    \label{fig:6}
\end{figure}
Also, from Eq~\ref{eq:I}, it therefore necessitates that 
\begin{equation}
    \acute{\xi}=-2\lambda^2 n_{0}
\end{equation}
which identifies $n_{0}$ as the energy parameter of this case. The energy level shift is again $\Delta\xi=\acute{\xi}_{k}-\xi_{k}=\frac{qB}{2}m_{\phi}$, directly proportional to the linear magnetic term. Therefore, Eq~\ref{eq:N} gives
\begin{equation}
\begin{aligned}
\frac{1}{2\lambda^2}
\left(H-\acute{\xi}\right)\phi_n(r)
&=
A_n x^{\alpha} e^{-\beta x}
\biggl\{
\frac{4n+4\alpha+1}{4}
+n_0
\\
&\quad
-\frac{x}{4}
+n_1 x
\biggr\}
L_n^{\nu}(x).
\end{aligned}
\end{equation}
Using the recurrence relation and orthogonality properties of the Laguerre polynomial, the wave equation becomes
\begin{equation}
\begin{aligned}
\Bigl[(2n_{2}+1)(n+\mu)+n_{0}\Bigr]f_{n}
&=n_{2}\Bigl[\sqrt{n(n+\nu)}\,f_{n-1}
\\
&\quad
+\sqrt{(n+1)(n+\nu+1)}\,f_{n+1}\Bigr],
\end{aligned}
\label{eq:T3}
\end{equation}
where $n_{2}=n_{1}-\frac{1}{4}$ and $\mu=\frac{\nu+1}{2}=\frac{2l+3}{4}$, so that $\alpha+\frac{1}{4}=\mu$. As previously noted in Eq~\ref{eq:T2}, Eq~\ref{eq:T3} possesses the same tridiagonal structure as the Meixner--Pollaczek recurrence relation, but with the roles exchanged: $n_{2}$ is now the potential parameter and $n_{0}$ the energy parameter. The identification with (B8) gives
\begin{equation}
    \mu=\frac{\nu+1}{2},\qquad \cos\varphi=\frac{2n_{2}+1}{2n_{2}},\qquad z\sin\varphi=\frac{n_{0}}{2n_{2}},
\end{equation}
and now $\cos^{2}\varphi-1=n_{1}/n_{2}^{2}>0$ for every $\eta>0$, independently of the energy. The spectrum is therefore purely discrete, as expected for a confining potential, and the analytic continuation $\varphi\to i\vartheta$ gives the discrete Meixner family with an energy-independent parameter. Similarly, working with the bound state energy spectrum obtained from the pole in the amplitude of the asymptotics of the Meixner-Pollaczek polynomial, $\mu+iz=-k$, one obtains $n_{0}^{2}=4(k+\mu)^{2}n_{1}$ with $n_{0}<0$, so that the resultant energy spectrum of the system becomes
\begin{equation}
    \acute{\xi}_{k}=4\lambda^{2}\sqrt{n_{1}}\,(k+\mu)=\sqrt{2\eta}\left(2k+l+\frac{3}{2}\right),\qquad
    \xi_{k,l,m_{\phi}}=\acute{\xi}_{k}-\frac{qB}{2}m_{\phi},
\label{eq:O}
\end{equation}
for $k=0,1,2,3,\ldots$ The spectrum is independent of $\lambda$, and for $\lambda^{2}=\sqrt{2\eta}$ (i.e. $n_{2}=0$) the basis of Eq.~\ref{eq:H} is the exact eigenbasis.
\begin{figure}
    \centering
    \includegraphics[width=0.7\linewidth]{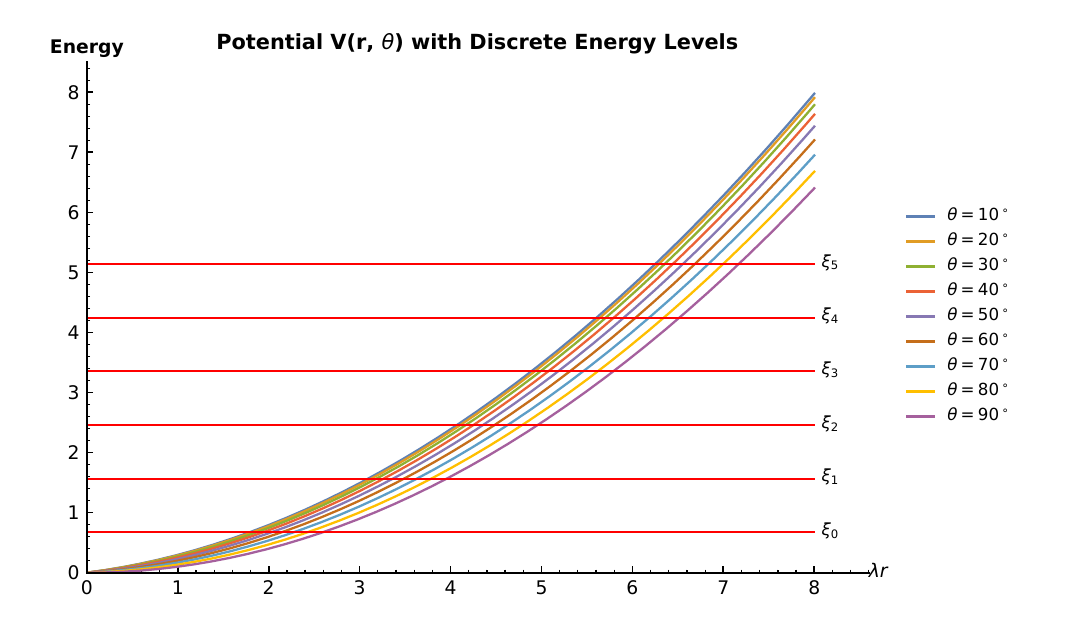}
    \caption{Graph of the potential function $V(r)$ Eq~\ref{eq:M} and the $\theta$-independent resultant energy spectrum $\acute{\xi}_{k}=\sqrt{2\eta}\,(2k+l+\frac{3}{2})$ Eq~\ref{eq:O}, $k=0,\ldots,5$. Same parameter values as figure 6 ($n_{2}=-0.2$), with $l=0$ and $\nu=l+\frac{1}{2}$. The vertical axis is in a dimensionless scaled energy unit, and the horizontal axis is in a dimensionless radial coordinate unit.}
    \label{fig:7}
\end{figure}
In this case the expansion coefficients are $f_{n}(\acute{\xi}_{k})\propto s^{n}M_{n}^{\mu}(k;\beta)$ with
\begin{equation}
    \beta=w^{2},\qquad w=\frac{\sqrt{2\eta}-\lambda^{2}}{\sqrt{2\eta}+\lambda^{2}},\qquad s=\operatorname{sgn}(w),
\end{equation}
where $\beta$ is now independent of $k$, so that the Meixner orthogonality applies directly. Since the basis of Eq.~\ref{eq:H} is orthonormal for $x=\lambda^{2}r^{2}$, the bound state wavefunction takes the form of Eq.~\ref{eq:J} with the normalization $\sqrt{\rho_{k}(\beta)}$,
\begin{equation}
    \psi\left(\acute{\xi}_{k},r\right)=\sqrt{\rho_{k}(\beta)}\sum_{n=0}^{\infty}s^{n}M_{n}^{\mu}(k;\beta)\,\phi_{n}(\lambda r),
\label{eq:J2}
\end{equation}
which sums to the closed form
\begin{equation}
    \psi_{k}(r)=N_{k}\,r^{l+1}e^{-\omega r^{2}/2}L_{k}^{l+\frac{1}{2}}(\omega r^{2}),\qquad \omega=\sqrt{2\eta},\qquad N_{k}^{2}=\frac{2\,\omega^{l+\frac{3}{2}}\,k!}{\Gamma\left(k+l+\frac{3}{2}\right)},
\end{equation}
as shown in Fig.~\ref{fig:8} and Fig.~\ref{fig:9}. 


\section{TRA Solution of the Effective Radial Schr\"{o}dinger Equation in Jacobi Basis}
\label{jacobi}
\begin{figure}
    \centering
    \includegraphics[width=0.7\linewidth]{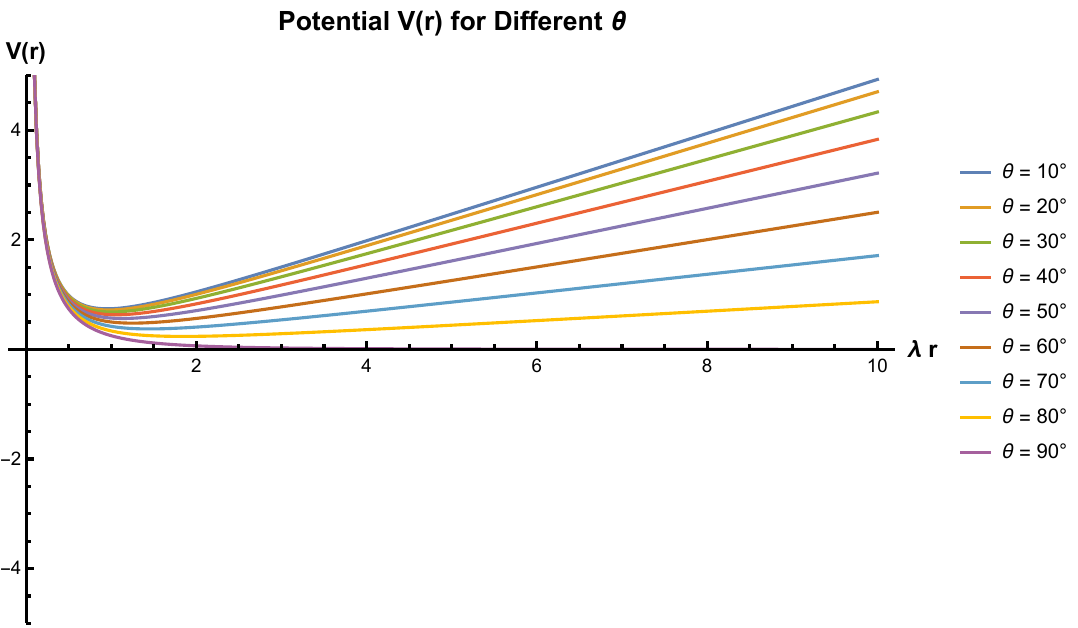}
    \caption{Graph of the potential function Eq~\ref{eq:40} where $r=\frac{\lambda(1+x)}{(1-x)}$ with $\lambda =1$, $E=1$,$q=0.5$, $C=\cos\theta$, $a_{1}=1$, and $a_{0}=2$. The vertical axis is in a dimensionless scaled energy unit, and the horizontal axis is in a dimensionless radial coordinate unit.}
    \label{fig:11}
\end{figure}

Here, the basis element takes the form 
\begin{equation}
    \phi_{n}=A_{n}\left( 1-x \right)^{\alpha}\left( 1+x \right)^{\beta}P^{(\mu,\nu)}_{n}(x)
\end{equation}
where $A_{n}=\sqrt{\frac{(2n+\mu+\nu+1)\Gamma(n+1)\Gamma(n+\mu+\nu+1)}{2^{\mu+\nu+1}\Gamma\left(n+\nu+1 \right)\Gamma\left( n+\mu+1 \right)}}$, $\alpha,\beta\ge 0$, $\mu,\nu>-1$, and $P^{(\mu,\nu)}_{n}(x)$ is a Jacobi polynomial. The transformation of the coordinate space takes the form $x=\frac{r-\lambda}{r+\lambda}$ such that $x\in \left[ -1,1 \right]$ for $r\in [0,\infty)$. Using this basis in Eq~\ref{eq:G}, with differential properties of the Jacobi basis (See Appendix C), we have 
\begin{equation}
\label{eq:P}
\begin{aligned}
-8\lambda^2
\left(H-\acute{\xi}\right)\phi_n(r)
&=
A_n(1-x)^{\alpha+3}(1+x)^{\beta-1}
\Biggl\{
(1-x^2)\frac{d^2}{dx^2}
\\
&\quad
+\Bigl[
2\beta(1-x)
+2(1+x)(\lambda-\alpha)
\Bigr]\frac{d}{dx}
\\
&\quad
+\frac{\beta(\beta-1)(1-x)}{1+x}
+\frac{\alpha(\alpha-1)(1+x)}{1-x}
\\
&\quad
-2\alpha\beta
+2\lambda\beta
-\frac{2\lambda\alpha(1+x)}{1-x}
\\
&\quad
-\frac{4l(l+1)}{(1+x)(1-x)}
\\
&\quad
-\frac{8\lambda^2(1+x)}
{(1-x)^3}
\bigl[\acute{V}(x)-\acute{\xi}\bigr]
\Biggr\}
P_n^{(\mu,\nu)}(x).
\end{aligned}
\end{equation}
Using the differential properties of the Jacobi polynomial, with the basis parameter taken as $\mu=2\alpha-2a-1$, $\nu=2\beta-1$ such that $a=-\lambda$, which demands $\lambda=1$ and $\beta(\beta-1)=l(l+1)$, then we have 
\begin{equation}
\label{eq:R}
\begin{aligned}
8\lambda^2
\left(H-\acute{\xi}\right)\phi_n(r)
&=
A_n(1-x)^{\alpha+3}(1+x)^{\beta-1}
\Biggl\{
n(n+\mu+\nu+1)
\\
&\quad
+\frac{\mu+\nu}{2}
\left(\frac{\mu+\nu}{2}-3\right)
\\
&\quad
+\frac{8\lambda^2(1+x)}
{(1-x)^3}
\bigl[\acute{V}(x)-\acute{\xi}\bigr]
\Biggr\}
P_n^{(\mu,\nu)}(x).
\end{aligned}
\end{equation}
One can see that the definitions of the basis parameters ensure the orthogonality of the Jacobi polynomial as
\begin{equation}
\label{eq:matrix_element}
\begin{aligned}
\langle \phi_n(r)|F(x)|\phi_m(r)\rangle
&=
A_nA_m
\int_{-1}^{1}
(1-x)^{2\alpha+3}
(1+x)^{2\beta-1}
F(x)
\\
&\quad\times
P_n^{(\mu,\nu)}(x)
P_m^{(\mu,\nu)}(x)
\frac{-2\lambda}{(1-x)^2}\,dx
\\[6pt]
&=
A_nA_m
\int_{-1}^{1}
(1-x)^{\mu}
(1+x)^{\nu}
\bigl[2aF(x)\bigr]
\\
&\quad\times
P_n^{(\mu,\nu)}(x)
P_m^{(\mu,\nu)}(x)\,dx .
\end{aligned}
\end{equation}
where $F(x)=n(n+\mu+\nu+1) +\left(\frac{\mu+\nu}{2}\right)
\left(\frac{\mu+\nu}{2}-3\right)+\frac{(1+x)\,8\lambda^2
\left[\acute{V}(x)-\acute{\xi}\right]}
{(1-x)^3}$. To ensure tridiagonality from Eq~\ref{eq:R}, we take similiar potential parameterization as before
\begin{equation}
    \frac{(1+x)\,8\lambda^2
\left[V(x)-\acute{\xi}\right]}
{(1-x)^3}=qEC\frac{(1+x)^{2}8\lambda^3}{(1-x)^4}+a_{1}x+a_{0}
\label{eq:Q}
\end{equation}
which gives the potential function as
\begin{equation}
    V(x)=A\frac{(1+x)}{(1-x)}+\frac{(a_{1}x+a_{0})(1-x)^{3}}{(1+x)8\lambda^2}
\label{eq:40}
\end{equation}
where $A=qEC\lambda$. Eq~\ref{eq:40} is a potential function $V(x)$ on $-1< x < 1$ with singularity at $\pm 1$. As $x\to 1^{-1}$, the potential takes the form $V(x)\approx \frac{2A}{1-x}$, while as $x\to -1^{+}$, $V(x)\approx \frac{a_{0}-a_{1}}{\lambda(1+x)}$.
So if $a_{0} > a_{1}$ and $A >0$, then the potential tends to $ + \infty $ at both boundaries, so it possesses at least one interior minimum and defines a confining well on $ (-1,1) $. In that case, the associated quantum problem has a discrete spectrum and supports bound states. For $0<\theta<90^{\circ}$, since $\cos\theta>0$, this bound-state support follows whenever $ A > 0 $. This type of potential arises naturally in effective one-dimensional models of quantum confinement where external fields and boundary effects interplay. It can be used to describe charged particles subject to non-uniform electric fields or anisotropic environments, where the interaction strength can be continuously tuned via the angle parameter. The presence of singular behavior at the boundaries makes it a useful prototype for studying particles confined within finite domains with strongly varying or asymmetric walls, such as quantum wells or nanostructures with edge effects. From a theoretical perspective, it provides a valuable example in mathematical quantum mechanics for analyzing spectral properties, particularly the emergence and disappearance of bound states as parameters vary, and the role of boundary conditions in systems with singular endpoints. More broadly, such potentials can serve as simplified models of low-dimensional condensed-matter or atomic systems, where confinement, external fields, and tunability combine to control localization and energy spectra. The absence of a closed-form expression for the
 resultant energy spectrum $\acute{\xi}$ does not mean that it is absent in this case; it means that $\acute{\xi}$ is obtained directly from the tridiagonal or generalized eigenvalue problem, and the physical spectrum then follows from
\begin{equation}
    \xi=\acute{\xi}-\frac{qBm_{\phi}}{2}
\end{equation}
once appropriate parameter values are substituted for the linear magnetic term.
\begin{figure}
    \centering
    \includegraphics[width=0.7\linewidth]{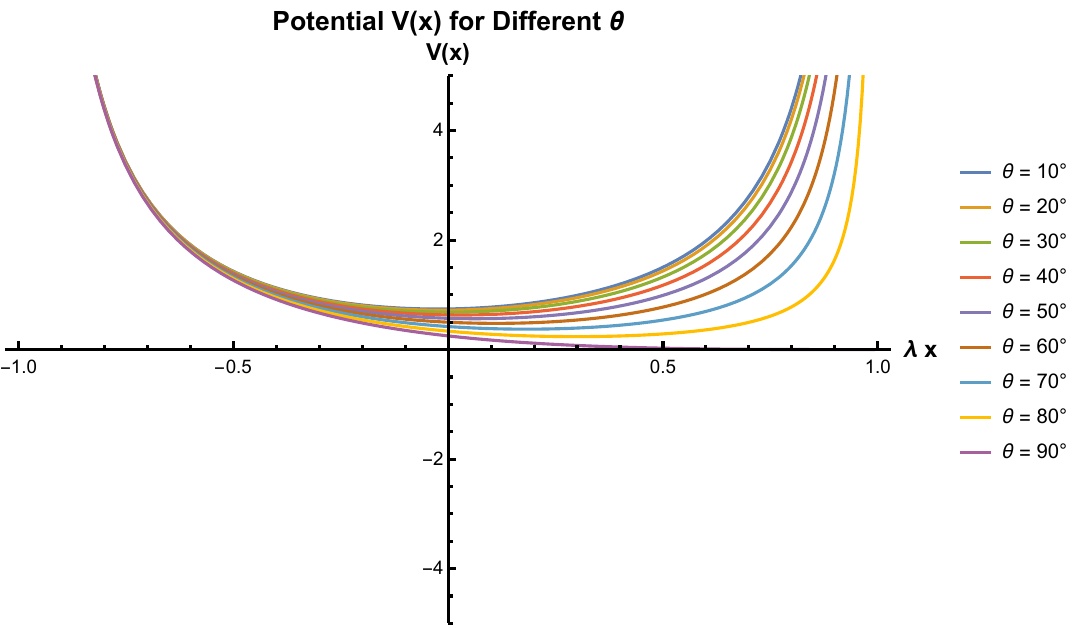}
    \caption{Graph of the potential function, \Eqref{eq:40}, with $\lambda =1$, $E=1$, $q=0.5$, $C=\cos\theta$, $a_{1}=1$, and $a_{0}=2$. The vertical axis is in a dimensionless scaled energy unit, and the horizontal axis is in a dimensionless $x$ coordinate unit.}
    \label{fig:10}
\end{figure}
On the other hand, if we return to the previous space with $r=\frac{\lambda(1+x)}{(1-x)}$, the potential looks graphically as shown in Fig.~\ref{fig:11}. 
So from Eq~\ref{eq:R}, we have
\begin{equation}
\label{eq:multiline_hamiltonian}
\begin{aligned}
8\lambda^2
\left(H-\acute{\xi}\right)\phi_n(r)
&=
A_n(1-x)^{\alpha+3}(1+x)^{\beta-1}
2a\Biggl\{
n(n+\mu+\nu+1)
\\
&\quad
+\frac{\mu+\nu}{2}
\left(\frac{\mu+\nu}{2}-3\right)
+a_1x+a_0
\Biggr\}
P_n^{(\mu,\nu)}(x).
\end{aligned}
\end{equation}
Finally, using the recurrence relation and orthogonality properties of the Jacobi polynomial, the wave equation becomes  
\begin{equation}
\label{eq:Z}
    -\frac{4\acute{\xi}}{\lambda^{2}}f_{n}=\acute{B}_{n}f_{n}+2aa_{1}\left[ D_{n-1}f_{n-1}+D_{n}f_{n+1}\right]
\end{equation} 
where $a_{0}=\frac{-2\acute{\xi}}{\lambda^{2}}$, $\acute{B}_{n}=\left\{2na(n+\mu+\nu+1) +\frac{a(\mu+\nu)(\mu+\nu-6)}{2}+\frac{2a(\nu^2-\mu^2)a_{1}}{(2n+\mu+\nu)(2n+\mu+\nu+2)}\right\}$, and 
\[
D_{n}
=
\frac{2}{2n+\mu+\nu+2}
\sqrt{
\frac{
(n+1)(n+\mu+1)(n+\nu+1)(n+\mu+\nu+1)
}{
(2n+\mu+\nu+1)(2n+\mu+\nu+3)
}
}.
\]
It should be noted that the parameters $a_{0}$ and $a_{1}$ in Eq~\ref{eq:Z} now play different roles. They are energy polynomial parameters that belong to the three-term recursion relation dervied from the wave operator. This  should not be taken as the potential function parameters. It is possible to replace these parameters with another hence this clarification. The above three-term recurrence relation has a similar structure
\begin{equation}
    zf_{n}=(B_{n}+\sigma C_{n})f_{n}+\sigma\left[ D_{n-1}f_{n-1}+D_{n}f_{n+1} \right]
\label{eq:S}
\end{equation}
where $z=\frac{-4\acute{\xi}}{\lambda^2}$, $B_{n}=\left\{ 2na(n+\mu+\nu+1)+\frac{a(\mu+\nu)(\mu+\nu-6)}{2} \right\}$, $C_{n}=\frac{(\nu^2 -\mu^2)}{(2n+\mu+\nu)(2n+\mu+\nu+2)}$, and $\sigma=2aa_{1}$.
The three terms in the recursion relation above do not correspond to any well-known orthogonal polynomial in the mathematical literature, which is one of the impressive impacts of the tridiagonal representation approach. The analytic properties, its weight function, generating functions, orthogonality, zeros, etc., are yet to be determined. A similar orthogonal polynomial has been mentioned in Appendix E \cite{Alhaidari2017}. However, we give a summary of this polynomial as shown in Appendix C. By Favard theorem's \cite{Ismail2005}, from Eq~\ref{eq:S}, $f_{n}(\acute{\xi})=f_{0}(\acute{\xi})Q_{n}(\acute{\xi})$, such that $f_{0}(\acute{\xi})=1$, then one can denote $Q_{n}(\acute{\xi})=\tilde{H}_{n}^{(\mu,\nu)}(z;\sigma)$. This is an orthogonal polynomial with a similar structure to that shown in Appendix C. The physical problems associated with this polynomial have only a continuous spectrum. However, from the study of these problems, the orthogonal polynomial has two discrete forms, one with an infinite spectrum denoted by $\tilde{h}_{n}^{(\mu,\nu)}(z_{k};\sigma)$ and the other with finite spectrum denoted by $\tilde{k}_{n}^{(\mu,\nu)}(z_{k};\sigma)$ with parameter choice dependent upon $\nu$, $\mu$, and $\sigma$. These polynomials are expected to be the expansion coefficients (with their weight function) of the bound state wavefunction. Diagonalizing the recursion relation by taking $a_{1}=0$ which implies that $\sigma=0$, then the bound energy spectrum becomes 
\begin{equation}
    \acute{\xi}_{k}=\frac{-\lambda^{2}}{8}\left[ 4ka(k+\mu+\nu+1)+a\left( \mu+\nu \right) \left( \mu+\nu-6 \right)\right]
\end{equation}
where $k = 0,1,2,3,....$ .To have estimated values for the energy spectrum, we resort to a numerical procedure. The first kind polynomial solution of the Eq~\ref{eq:S} has $\tilde{H}_{0}^{(\mu,\nu)}(z;\sigma)=1$ and $\tilde{H}_{1}^{(\mu,\nu)}(z;\sigma)=\frac{\left(\frac{-4\acute{\xi}}{\lambda^2} -\acute{B}_{0}\right)}{2a_{1}a D_{0}}H_{0}^{(\mu,\nu)}(z;\sigma)$ where $z_{0}=\frac{-4\acute{\xi}}{\lambda^2}$. This relation can be used recursively for $n = 0, 1, 2, 3,... $. However, we first obtain the Hamiltonian matrix from the wave operator by taking $\acute{\xi} = 0$ on the left-hand side of Eq~\ref{eq:S}. Therefore, we show how the energy spectrum can be obtained from the generalized eigenvalues $\left\{ \acute{\xi} \right\}$  of the matrix equation 
\begin{equation}
    \sum_{m} H_{n,m} f_{m} \approx \acute{\xi}\sum_{m}\Omega_{n,m}\,f_{m}
\end{equation}
where $H_{n,m}=\acute{B}_{n}\delta_{n,m}+2\sigma\left( D_{n-1}\delta_{n,m+1}+D_{n}\delta_{n,m-1}\right)$; and for easy numerical computation based on the complication of the three term recursion relation of the Jacobi polynomial; we approximately take $\Omega_{n,m}\approx \left( 1+4C_{n}\right)\delta_{n,m}+4D_{n-1}\delta_{n,m+1}+4D_{n}\delta_{n,m-1}$. The overlap matrix $\Omega_{n,m}$ was approximated (second-order accurate) by retaining only the tridiagonal contributions generated by the Jacobi three-term recurrence relation. The neglected remainder contains higher-band couplings $|n-m|\ge 2$. Since $C_{n}=O(n^{-2})$ and $D_{n}\to 1/2$ as $n\to \infty$, then the truncation error satisfies $||\Omega-\Omega_{T}||=O(N^{-2})$. Consequently, the computed generalized eigenvalues converge as $O(N^{-2})$ increasing basis size $N$ that is $|\acute{\xi}_{k}-\hat{\acute{\xi}}_{k}|=O(N^{-2})$.  Table 1 presents a few lists of the energy spectrum for a given set of polynomial parameter values and a given basis size. The accuracy would have improved if the basis elements were orthonormal in this case, or if we had an exact energy spectrum formula from the asymptotic expansion of this orthogonal polynomial, yet convergence still exists. However, we show only the significant decimal digits of the relative change with increasing basis size $N$.

\begin{table*}[htbp]
\centering
\caption{List of few resultant eigenvalues $\acute{\xi}_k$ with respect to the basis size $N$ with $\nu=-0.8$, $\mu=1.2$, $a=-1.0$, and $a_{1}=-2.0$.}
\label{tab:eigenvalues}
\renewcommand{\arraystretch}{1.3}
\setlength{\tabcolsep}{10pt}
\begin{tabular}{c c c c c c c}
\hline\hline
$k$ &
$\acute{\xi}_k~(N=10)$ &
$\acute{\xi}_k~(N=15)$ &
$\acute{\xi}_k~(N=20)$ &
$\acute{\xi}_k~(N=30)$ &
$\acute{\xi}_k~(N=40)$ &
$\acute{\xi}_k~(N=50)$ \\
\hline
0 & 0.64557618 & 0.64557618 & 0.64557618 & 0.64557618 & 0.64557618 & 0.64557618 \\
1 & 3.9930397  & 3.9930397  & 3.9930397  & 3.9930397  & 3.9930397  & 3.9930397  \\
2 & 9.4289776  & 9.4286019  & 9.4286019  & 9.4286019  & 9.4286019  & 9.4286019  \\
3 & 17.240146  & 16.955812  & 16.955812  & 16.955812  & 16.955812  & 16.955812  \\
4 & 37.065242  & 26.575426  & 26.574204  & 26.574204  & 26.574204  & 26.574204  \\
\hline\hline
\end{tabular}
\end{table*}
For numerical reproducibility, the Jacobi-basis calculation should be interpreted as a Galerkin-type truncation of the generalized eigenvalue problem. In the present version, the basis size $N$ is varied, and the low-lying eigenvalues are reported only after stabilization with respect to $N$. The tabulated convergence is therefore part of the validation in the Jacobi case. 

\section{Conclusion and Discussion}
We have formulated and analyzed effective radial Schr\"{o}dinger models motivated by weak-field approximation. The construction relies on the angular projection replacement by $C=\cos\theta$ and should therefore be interpreted as an effective radial approximation, not as an exact solution of the full nonseparable Stark problem. This distinction is essential for the physical interpretation of the spectra: the reported energies are eigenvalues of the projected radial models, followed by the diagonal energy shift. They are valid only within the stated weak-field and orientation-localization assumptions.

Within this modeling framework, the Tridiagonal Representation Approach provides a systematic algebraic route to recurrence relations for expansion coefficients in Laguerre and Jacobi bases. The Laguerre construction identifies two solvable field-modified central potentials, namely a linear-plus-inverse potential and a linear-plus-quadratic potential. In both cases, square-integrability and confinement require explicit restrictions on the parameters, and the coefficient recurrences can be connected with Meixner--Pollaczek and Meixner polynomial type structures. The resulting resultant spectra are of Coulomb-like and oscillator-like form, $\acute{\xi}_{k}=-b^{2}/[2(k+l+1)^{2}]$ and $\acute{\xi}_{k}=\sqrt{2\eta}\,(2k+l+\frac{3}{2})$, independent of the basis scale, the orientation, and the electric field; the latter two enter through the shape of the effective central potential $V(r)$. The Jacobi construction maps the radial half-line onto a finite interval and yields a singular, boundary-confined effective potential. Its recurrence relation is not reducible, in the form obtained here, to a standard classical family; consequently, the spectrum is obtained through a generalized eigenvalue problem. All the energy spectra obtained in both cases are coupled linear magnetic term to give result energy spectrum; therefore obtaining the energy level shift which are directly proportional to the linear magentic term. It is should be noted that the effective central potential function obtained here plotted for several values of $\theta$ within $0\le  \theta \le 90$ as can be seen in the plots.

So far, the effective field modified central potentials considered are within the boundaries $\left( 0,\infty\right)$; other possible boundaries are $\left( -\infty,\infty\right)$ and the finite interval $\left( -a,a\right)$. Future work should include more feasible domains for the effective-field-modified central-radial potentials that provide exact overlap-matrix implementations for the Jacobi, Laguerre, and other bases, such as the Bessel. 

The most direct experimental benchmark for the present effective radial model is a single trapped ion in a Penning architecture, where static electric and magnetic fields confine a charged particle. Recent work has demonstrated full quantum control of a single ion in a micro-fabricated Penning trap operating at 3 T, thereby establishing a realistic platform for field-dependent spectral measurements \cite{Bonus F 2025}. In parallel, trapped-ion electrometry using a static magnetic-field gradient has shown that electric-field-induced displacements can be detected with very high sensitivity over a broad bandwidth 
\cite{Duprez2024}. These developments make the harmonic-plus-linear branch of the present model particularly suitable for experimental validation through measurements of equilibrium-position shifts, Zeeman-shifted low-lying levels, and field-controlled wavefunction localization. As a complementary condensed-matter platform, recent spectroscopy of gate-defined bilayer graphene quantum dots has resolved discrete excited states under perpendicular and parallel magnetic fields 
\cite{Denisov AO 2025,Paul AS 2024}, while nanofabricated $Cu2O$ structures provide an attractive excitonic analog for the Coulomb-like branch. Looking forward, this framework provides a strong foundation for further generalization to higher-dimensional systems. In particular, extending the model to two-dimensional geometries opens promising avenues for applications in condensed matter physics, where external fields, confinement, and reduced dimensionality play a central role in phenomena such as quantum wells, quantum dots, and surface states. Overall, the approach developed in this work offers a flexible and powerful tool for exploring field-modified quantum systems across a wide range of physical settings.

\section*{Acknowledgements}
This research was funded by Khalifa University of Science and Technology through Project ID KU-INT-RIG-2024-8474000739.

Furthermore, T. J. Osunmusanmi acknowledges the support of the Saudi Centre for Theoretical Physics, Dhahran, Saudi Arabia.

\section*{Data Availability}
The data supporting the findings of this study are included in the article. Numerical input parameters needed to reproduce the plotted spectra are reported in the figure captions and in the text; additional scripts can be provided by the corresponding author upon reasonable request.

\section*{Competing interests}
The authors declare no competing interests.

\appendix
\renewcommand{\theequation}{A\arabic{equation}}
\renewcommand{\theHequation}{A\arabic{equation}}
\setcounter{equation}{0}
\section*{Appendix A: Review of Tridiagonal Representation Approach}
The Tridiagonal Representation Approach can be explained as follows. Given a differential equation
\begin{equation}
    D\psi\left(x\right)=\left[ p(x)\frac{d^2}{dx^2} +q(x)\frac{d}{dx}+g(x)\right]\psi\left( x \right)=0
\end{equation}
for some proper choice of functions $p(x)$, $q(x)$, and $g(x)$. The solution is written as the following bounded point-wise convergent series
\begin{equation}
    \psi (x) =\sum _{n=0}^{\infty }f_{n}\phi_{n}\left(y\right)
\end{equation}
where $y=y(x)$ is an independent variable transformation, $\left\{ \phi_{n}(y)\right\}$ is a complete set of square integrable functions, and $\left\{ f_{n}\right\}$ are the expansion coefficients. The basis is required to carry a tridiagonal matrix representation for the differential operator $D$. That is, the action of $D$ on the basis element should read 
\begin{equation}
    D\phi_{n}\left( y \right)=\omega\left(y\right)\left[ a_{n} \phi_{n}(y)+b_{n-1}\phi_{n-1}(y)+c_{n}\phi_{n+1}(y)\right]
\end{equation}
where $\omega(y)$ is a node-less entire function and $\left\{ a_{n},b_{n},c_{n} \right\}$ are constant coefficients. Hence, the differential equation, $D\psi (x) =0$, will become three term recursion relation for the expansion coefficients $\left\{ f_{n}\right\}$ that reads
\begin{equation}
    a_{n}F_{n}+c_{n-1}F_{n-1}+b_{n}F_{n+1}
\end{equation}
where we write $f_{n}=f_{0}F_{n}$, making $F_{0} =1$. Thus, the solution to the differential equation $(A1)$ changes to an algebraic solution of the discrete relation $(A4)$. Now, the set $\left\{ f_{n}\right\}$ contains all the properties of the solution $\psi (x)$  as mentioned earlier. There are two solution scenarios for $(A4)$
\begin{itemize}
    \item $b_{n} c_{n} > 0$, for all $n=0,1,2,...$
    \item $b_{n} c_{n} > 0$, only for $n=0,1,2,...N$
\end{itemize}
In the first case, $\left\{F_{n}\right\}_{n=0}^{\infty}$ will become an infinite set of orthogonal polynomials that has the following generalized orthogonality (continuous and discrete) 
\begin{equation}
    \int_{z_{-}}^{z^{+}}\rho(z)F_{n}(z)F_{m}(z)dz+\sum_{k=0}^{K}\zeta(z_{k})F_{n}(z_{k})F_{m}(z_{k})=\xi_{n}\delta_{n,m}
\end{equation}
where $z$ is some proper function of the differential equation parameters with $a_{n}=u_{n}-zh_{n}$ such that $[u_{n},h_{n},c_{n},b_{n}]$ are independent of $z$. Moreover $\xi_{n}>0$, $K$ is finite or infinite, and $\rho(z)$ is proportional to $\left[f_{0}\right]^{2}$. Obviously, the basis set $\left\{\phi_{n}(y)\right\}$  must also be infinite in this case. In some cases, the solution is either purely continuous (no sum in the orthogonality) or purely discrete (no integral in the orthogonality). For the second case, $\left\{F_{n} \right\}_{n=0}^{N}$ becomes a finite set of orthogonal polynomials, and the basis set $\left\{\phi_{n}(y)\right\}$ must also be finite.

\section*{Appendix B: Orthogonal Polynomial in the Laguerre Basis}
\label{appB}
\renewcommand{\theequation}{B\arabic{equation}}
\renewcommand{\theHequation}{B\arabic{equation}}
\setcounter{equation}{0}

The Laguerre basis is given as 
\begin{equation}
    \phi_{n}(r)=A_{n}y^{\alpha}e^{-\beta y}L_{n}^{\nu}(y)
\end{equation}
where $\alpha$ ,$\beta$, $\nu$ are real parameters and $\alpha, \beta \ge 0$ to ensure convergence of the Laguerre polynomial $L_{n}^{\nu}(y)$ compatible with boundary condition with the new $y$  span semi -- infinite interval $[0,\infty]$. $A_{n}$ is chosen based on the derivative of the space transformation in order to ensure the orthogonal condition of the Laguerre polynomial is satisfied. The Laguerre polynomial also satisfies the following properties: 
\begin{equation}
\begin{aligned}
yL_n^{\nu}(y)
&=
(2n+\nu+1)L_n^{\nu}(y)
\\
&\quad
-(n+\nu)L_{n-1}^{\nu}(y)
-(n+1)L_{n+1}^{\nu}(y).
\end{aligned}
\end{equation}
\begin{equation}
    L_{n}^{\nu}(y)=\frac{\Gamma(n+\nu+1)}{\Gamma(n+1)\Gamma(\nu+1)}{}_1F_1(-n;\nu+1;y)
\end{equation}
\begin{equation}
    \left[ y\frac{d^2}{dx^2} +\left( \nu +1-y \right)\frac{d}{dy}+n\right]L_{n}^{\nu}(y)
\end{equation}
\begin{equation}
    y\frac{d}{dy}L_{n}^{\nu}(y)=nL_{n}^{\nu}(y)-\left( n+v \right)L_{n-1}^{\nu}(y)
\end{equation}
\begin{equation}
    \int_{0}^{\infty}y^{\nu}e^{-y}L_{n}^{\nu}(y)L_{m}^{\nu}(y)dy=\frac{\Gamma(n+\nu+1)}{\Gamma(n+1)}\delta_{n,m}
\end{equation}
When using the Laguerre bases, the wave equation often yields a three-term recursion relation for the expansion coefficients of the continuous wavefunction. These are identified with either the Meixner-- Pollaczek polynomial or the continuous dual Hahn polynomial. Here we give some of the properties of the Meixner- Pollaczek polynomial and its discrete forms. The discrete form of this polynomial is used as the expansion coefficients for the bound states' wavefunction. The normalized version for the Meixner - Pollaczek polynomials ensure that $\int_{z_{-}}^{z_{+}}\rho (z)P_{n}(z)P_{m}(z)dz=\delta_{n,m}$ and $P_{0} = 1$, where $\rho (z)$ is the normalized weight function. However, the discrete form satisfies the orthogonal $\sum_{k_{-}}^{k_{+}}\omega(k)Q_{n}(k)Q_{m}(k)=\delta_{n,m}$, where $Q_{n}(k)$ is the normalized discrete version of $P_{n}(z)$ and $\omega(k)$ is the discrete weight function.  The Meixner -- Pollazek normalized version is defined as 
\begin{equation}
\begin{aligned}
P_n^{\mu}(z,\varphi)
&=
\sqrt{
\frac{\Gamma(n+2\mu)}
     {\Gamma(2\mu)\Gamma(n+1)}
}
\,e^{in\varphi}
\\
&\quad\times
{}_2F_1\!\left(
\begin{array}{c}
-n,\;\mu+iz \\
2\mu
\end{array}
\,;\,
1-e^{-2i\varphi}
\right).
\end{aligned}
\end{equation}
where $z\in \left[ -\infty ,\infty \right]$, $\mu >0 $, and $0 < \varphi < \pi$. This is a polynomial with a continuous spectrum. It satisfies the three-term recursion relation 
\begin{equation}
\label{eq:recursion_relation}
\begin{aligned}
(z\sin\varphi)\,P_n^{\mu}(z,\varphi)
&=
-(n+\mu)\cos\varphi\,
P_n^{\mu}(z,\varphi)
\\
&\quad
+\frac{1}{2}\sqrt{n(n+2\mu-1)}
\,P_{n-1}^{\mu}(z,\varphi)
\\
&\quad
+\frac{1}{2}\sqrt{(n+1)(n+2\mu)}
\,P_{n+1}^{\mu}(z,\varphi).
\end{aligned}
\end{equation}
with the corresponding normalized weight function 
\begin{equation}
    \rho^{\mu}\left( z,\varphi \right)=\frac{1}{2\pi \Gamma(2\mu)}(2\sin\varphi)^{2\mu}e^{(2\varphi -\pi)z}|\Gamma(\mu+iz)|^{2}
\end{equation}
As $n\to \infty$ the polynomial reads 
\begin{equation}
\begin{aligned}
P_n^{\mu}(z,\varphi)
&\approx
\frac{
2n^{-1/2}
e^{(\pi/2-\varphi)z}
}{
(2\sin\varphi)^{\mu}
|\Gamma(\mu+iz)|
}
\\
&\quad\times
\cos\Bigl[
n\varphi
+\arg\Gamma(\mu+iz)
-\frac{\mu\pi}{2}
\\
&\qquad\qquad
-2\ln(2n\sin\varphi)
\Bigr].
\end{aligned}
\end{equation}
The scattering phase shift $(mod$ $\pi/2)$ reads as 
\begin{equation}
    \delta^{\mu}(z)=arg\Gamma\left( \mu +iz \right) -\mu\pi/2
\end{equation}
Since the scattering amplitude in $(B10)$ vanishes at $\mu +iz =-k$, then the discrete spectrum formula associated with this polynomial is $z^{2}=-(k+\mu)^2$ which is infinite for positive $\mu$ and finite for negative $\mu$ as $k=0,1,2,..,N$ where $N$  is the largest integer less than or equal to $-\mu$. The generating function of Meixner -- Pollaczek is 
\begin{equation}
    \sum_{n=0}^{\infty}\tilde{P}_{n}^{\mu}(z,\varphi)t^{n}=(1-te^{i\varphi})^{-\mu+iz}(1-te^{-i\varphi})^{-\mu-iz}
\end{equation}
where $\tilde{P}_{n}^{\mu}(z,\varphi)=\sqrt{\frac{\Gamma(n+2\mu)}{\Gamma(2\mu)\Gamma(n+1)}}P_{n}^{\mu}(z,\varphi)$. The first normalized discrete form of this polynomial with infinite spectrum is called Meixner polynomial which is defined as 
\begin{equation}
    M_{n}^{\mu}(m,\beta)=\sqrt{\frac{\Gamma(n+2\mu)\beta^{n}}{\Gamma(2\mu)\Gamma(n+1)}}{}_2F_1\!\left(
\begin{array}{c}
-n,\; -m \\
2\mu
\end{array}
; 1-\beta^{-1}
\right)
\end{equation}
where $0 < \beta < 1$ and $\mu >0$. Its three-term recursion relation is
\begin{equation}
\begin{aligned}
(1-\beta)\,m\,M_n^{\mu}(m;\beta)
&=
\bigl[n(1+\beta)+2\mu\beta\bigr]
M_n^{\mu}(m;\beta)
\\
&\quad
-\sqrt{n(n+2\mu-1)\beta}\,
M_{n-1}^{\mu}(m;\beta)
\\
&\quad
-\sqrt{(n+1)(n+2\mu)\beta}\,
M_{n+1}^{\mu}(m;\beta).
\end{aligned}
\end{equation}
The normalized discrete weight function is $\rho_{m}^{\mu}(\beta)=\beta^{m}(1-\beta)^{2\mu}\frac{\Gamma(m+2\mu)}{\Gamma(2\mu)\Gamma(m+1)}$. Another form of this recursion relation can be obtained from (B8) by replacing $\theta\to i\theta$, which directly makes $\cos\theta\to \pm  \cosh\theta$ and $\sin\theta\to \sinh\theta$. The second normalized discrete form is the Krawtchouk polynomial (with finite spectrum), defined as 
\begin{equation}
    K_{n}^{\mu}(m;\beta)=\sqrt{\frac{\Gamma(N+1)\beta^{n}/(1-\beta)^n}{\Gamma(N-n+1)\Gamma(n+1)}}{}_2F_1\!\left(
\begin{array}{c}
-n,\; -m \\
N
\end{array}
; \beta^{-1}
\right)
\end{equation}
where $0< \beta < 1$ and $n,m=0,1,2,..,N$. Also, its recurrence relation is 
\begin{equation}
\begin{aligned}
m K_{n}^{\mu}(m;\beta)
&= \left[n(1-\beta)+\beta(N-n)\right] K_{n}^{\mu}(m;\beta) \\
&\quad - \sqrt{\beta(1-\beta)n(N-n+1)}\, K_{n-1}^{\mu}(m;\beta) \\
&\quad - \sqrt{\beta(1-\beta)(n+1)(N-n)}\, K_{n+1}^{\mu}(m;\beta).
\end{aligned}
\end{equation}
and its normalized discrete weight function is $\rho_{m}^{N}(\beta)=\beta^{m}(1-\beta)^{N-m}\frac{\Gamma(N+1)}{\Gamma(m+1)\Gamma(N-m+1)}$.

\section*{Appendix C: Orthogonal Polynomial in the Jacobi Basis}
\label{appC}
\renewcommand{\theequation}{C\arabic{equation}}
\renewcommand{\theHequation}{C\arabic{equation}}
\setcounter{equation}{0}

The Jacobi basis element is given
\begin{equation}
    \phi(r)=A_{n}(1-y)^{\alpha}(1+y)^{\beta}P_{n}^{(\mu,\nu)}(y)
\end{equation}
where $\alpha,\beta\ge 0$, $\mu,\nu >-1$ to ensure convergence of the Jacobi polynomial $P_{n}^{(\mu,\nu)}(y)$ compatible with boundary condition with the new  $y$  span finite interval $[-1,1]$. $A_{n}$ is chosen based on the derivative of the space transformation in order to ensure the orthogonal condition of the Jacobi polynomial is satisfied. The Jacobi polynomial also satisfies the following properties:  
\begin{equation}
\begin{aligned}
P_n^{(\mu,\nu)}(y)
&=
\frac{\Gamma(n+\mu+1)}
     {\Gamma(n+1)\Gamma(\mu+1)}
\\
&\quad\times
{}_2F_1\!\left(
-n,\,
n+\mu+\nu+1;\,
\mu+1;\,
\frac{1-y}{2}
\right)
\\
&=
(-1)^n P_n^{(\mu,\nu)}(y).
\end{aligned}
\end{equation}
\begin{equation}
\begin{aligned}
\Biggl\{
&(1-y^2)\frac{d^2}{dy^2}
\\
&-\Bigl[(\mu+\nu+2)y+\mu-\nu\Bigr]
\frac{d}{dy}
\\
&+n(n+\mu+\nu+1)
\Biggr\}
P_n^{(\mu,\nu)}(y)
=0.
\end{aligned}
\end{equation}
\begin{equation}
\begin{aligned}
(1-y^2)\frac{d}{dy}P_n^{(\mu,\nu)}(y)
&=
-n\left(
y+\frac{\nu-\mu}{2n+\mu+\nu}
\right)
P_n^{(\mu,\nu)}(y)
\\
&\quad
+\frac{2(n+\mu)(n+\nu)}
       {2n+\mu+\nu}
P_{n-1}^{(\mu,\nu)}(y).
\end{aligned}
\end{equation}
\begin{equation}
\begin{aligned}
\int_{-1}^{1}
&(1-y)^{\mu}(1+y)^{\nu}
P_n^{(\mu,\nu)}(y)
P_m^{(\mu,\nu)}(y)\,dy
\\
&=
\frac{
2^{\mu+\nu+1}
\Gamma(n+\mu+1)
\Gamma(n+\nu+1)
}{
(2n+\mu+\nu+1)
\Gamma(n+1)
\Gamma(n+\mu+\nu+1)
}
\,\delta_{n,m}.
\end{aligned}
\end{equation}
\begin{equation}
\begin{aligned}
\left(\frac{1\pm y}{2}\right)
P_n^{(\mu,\nu)}(y)
&=
\frac{
2n(n+\mu+\nu+1)
}{
(2n+\mu+\nu)
(2n+\mu+\nu+2)
}
P_n^{(\mu,\nu)}(y)
\\
&\quad
+
\frac{
(\mu+\nu)
\left[
\frac{\mu+\nu}{2}
\pm\frac{\nu-\mu}{2}
+1
\right]
}{
(2n+\mu+\nu)
(2n+\mu+\nu+2)
}
P_n^{(\mu,\nu)}(y)
\\
&\quad
\pm
\frac{(n+\mu)(n+\nu)}
     {(2n+\mu+\nu)(2n+\mu+\nu+1)}
P_{n-1}^{(\mu,\nu)}(y)
\\
&\quad
\pm
\frac{(n+1)(n+\mu+\nu+1)}
     {(2n+\mu+\nu+1)(2n+\mu+\nu+2)}
P_{n+1}^{(\mu,\nu)}(y).
\end{aligned}
\end{equation}
\begin{equation}
\begin{aligned}
y\,P_{n}^{(\mu,\nu)}(y)
&= \frac{\nu^2-\mu^2}{(2n+\mu+\nu)(2n+\mu+\nu+2)}\,P_{n}^{(\mu,\nu)}(y) \\
&\quad + \frac{2(n+\mu)(n+\nu)}{(2n+\mu+\nu)(2n+\mu+\nu+1)}\,P_{n-1}^{(\mu,\nu)}(y) \\
&\quad + \frac{2(n+1)(n+\mu+\nu+1)}{(2n+\mu+\nu+1)(2n+\mu+\nu+2)}\,P_{n+1}^{(\mu,\nu)}(y).
\end{aligned}
\end{equation}
Similar to Laguerre bases, when using the Jacobi bases, the wave equation often leads to the three-term recursion relation for the expansion coefficient of the continuous wavefunction. However, these recursion relations and their associated orthogonal polynomials are not treated in the mathematical literature, and their analytic properties (weight function, generating function, orthogonality, differential property, spectrum formula, asymptotics, etc.) have yet to be derived. Here, we mention one of them (related to \Eqref{eq:S} and give some insights about them. More details about this new polynomial can be found in \cite{Alhaidari2017}. This polynomial is a three-parameter polynomial designated as
$H_{n}^{(\mu,\nu)}(z;\sigma)$. It is a three-term recursion relation for $n=0,1,2,...,$
\begin{equation}
\begin{aligned}
zH_n^{(\mu,\nu)}(z;\sigma)
&=
\left(B_n^2+\sigma C_n\right)
H_n^{(\mu,\nu)}(z;\sigma)
\\
&\quad
+\sigma\Bigl[
D_{n-1}H_{n-1}^{(\mu,\nu)}(z;\sigma)
\\
&\qquad
+D_nH_{n+1}^{(\mu,\nu)}(z;\sigma)
\Bigr].
\end{aligned}
\end{equation}
where $B_{n}=\left( n+\frac{\mu+\nu+1}{2} \right)$, $C_{n}=\frac{\nu^2-\mu^2}{(2n+\mu+\nu)(2n+\mu+\nu+2)}$, and $D_{n}=\frac{2}{(2n+\mu+\nu+2)}\sqrt{\frac{(n+1)(n+\mu+1)(n+\nu+1)(n+\mu+\nu+1)}{(2n+\mu+v+1)(2n+\mu+\nu+3)}}$. Alternatively we can have it has 
\begin{equation}
\begin{aligned}
z\tilde{H}_{n}^{(\mu,\nu)}(z;\sigma)
&= \left(B_{n}^{2}+\sigma C_{n}\right)
   \tilde{H}_{n}^{(\mu,\nu)}(z;\sigma) \\
&\quad + \frac{2\sigma(n+\mu)(n+\nu)}
{(2n+\mu+\nu)(2n+\mu+\nu+1)}
   \tilde{H}_{n-1}^{(\mu,\nu)}(z;\sigma) \\
&\quad + \frac{2\sigma(n+1)(n+\mu+\nu+1)}
{(2n+\mu+\nu+1)(2n+\mu+\nu+2)}
   \tilde{H}_{n+1}^{(\mu,\nu)}(z;\sigma).
\end{aligned}
\end{equation}
where $H_{n}^{(\mu,\nu)}(z;\sigma)=A_{n}\tilde{H}_{n}^{(\mu,\nu)}(z;\sigma)$ such that $A_{n}=\sqrt{\frac{2n+\mu+\nu+1}{\mu+\nu+1}\frac{n!(\mu+\nu+1)}{(\mu+1)_{n}(\nu+1)_{n}}}$. $H_{n}^{(\mu,\nu)}(z;\sigma)$ is the orthonormal version of $\tilde{H}_{n}^{(\mu,\nu)}(z;\sigma)$. The initial values of for first kind polynomial solution (equation C8) take the form $H_{0}^{(\mu,\nu)}(z;\sigma)=1$ and $H_{1}^{(\mu,\nu)}(z;\sigma)=(\sigma D_{0})^{-1}\times \left[ z-\sigma C_{0}-\frac{1}{4} (\mu+\nu+1)^2\right]$. This polynomial is also known as a dipole polynomial. It has the Jacobi polynomial as a special case $P_{n}^{(\mu,\nu)}(y)=\lim_{\sigma,z \to \infty} \tilde{H}_{0}^{(\mu,\nu)}(z;\sigma)$ such that $\lim_{\sigma,z \to \infty} \left( z/\sigma \right)=y\in \left[ -1,1 \right]$.

\end{document}